----------
X-Sun-Data-Type: default
X-Sun-Data-Name: cuo.tex
X-Sun-Content-Lines: 3643

\documentstyle[12pt]{article}

\renewcommand{\theequation}{\arabic{section}.\arabic{equation}}

\newcommand{\aSU}{\sum\nolimits}
\input {amssym.def}
\input {amssym.tex}
\everymath{\rm}
\everydisplay{\rm}
\small
\title{\bf Non Fermi-Liquid States and Pairing Instability of a General Model of Copper-Oxide Metals}
\author{C. M. Varma \\
Bell Laboratories, Lucent Technologies \\
Murray Hill, NJ 07974}
\begin{document}
\maketitle
\normalsize
\begin{abstract}
A model of copper-oxygen bonding and anti-bonding bands
with the most general two-body interactions allowable by
symmetry is considered.
The model has a continuous transition as a function of
hole-density $x$ and temperature T to a phase in which a current
circulates in each unit cell. This phase preserves the translational
symmetry of the lattice while breaking time-reversal invariance and the 
four-fold rotational symmetry. The product of time-reversal and four-fold 
rotation is preserved.  The circulating current phase terminates at a critical point at
$x=x_c$, $T=0$. In the quantum-critical region about this point the        logarithm of the frequency of
the current fluctuations scales with their momentum. The microscopic basis for the marginal Fermi-liquid 
phenemenology and the observed long wavelength transport
anomalies near $x=x_c$ are derived from such fluctuations.
The symmetry of the current fluctuations is such that 
the associated magnetic field fluctuations are absent at oxygen
sites and have the correct form to explain the anomalous
copper nuclear relaxation rate. Cross-overs to the Fermi-liquid phase on either side of $x_c$ and
the role of disorder are briefly considered.
The current fluctuations promote superconductive
instability with a propensity towards ``D-wave" symmetry 
or ``extended S-wave"symmetry
depending on details of the band-structure.
Several experiments are proposed to test the theory.
\end{abstract}
\setlength{\baselineskip}{1.5\baselineskip}
\newpage
\makeatletter
\def\subsection{\@startsection {subsection}{1}{\z@}{-3.5ex plus -1ex minus
 -.2ex}{2.3ex plus .2ex}{\normalsize\bf}}
\makeatother

\section{Introduction}
Besides their exceptionally high superconducting transition temperatures, Copper-Oxide (Cu-O)
based metals also have exceptional normal state properties.$^1$
Landau Fermi-liquid theory$^2$ and associated quasi particle concepts, which are a
foundation stone for much of our understanding of phenomena in condensed matter appear to
be inapplicable to their normal state.  The principal problem is the development of a
{\it consistent} theoretical framework in 
which the unusual metallic properties can be understood.
Moreover, it is {\it necessary}
to have a theory for the normal state to understand the mechanism
for the superconductive instability.

Every transport property in the normal state of Copper-Oxide metals
has a temperature dependence unlike other
metals.  For example, the electrical resistivity has a linear temperature dependence
down to $T_c$ for composition near the highest $T_c$ for any class of Cu-O compounds even
when that $T_c$ is as low as 10K.$^3$  
On the other hand, the equilibrium properties, such as specific
heat $C_v$ and magnetic susceptibility $\chi$
are consistent with the usual temperature dependence and are in fact only
about a factor of two enhanced over band structure calculations.$^4$  The copper-oxide metals
are thus qualitatively different from liquid $^3 He$ and heavy Fermion metals where strong
interactions produce strong quantitative renormalizations in both equilibrium and transport
properties without changing the asymptotic low temperature dependences and which are
properly called Fermi-liquids.

Recent re-examinations$^5$ of the foundations of Landau's Fermi-liquid theory have confirmed
the robustness of the theory for dimensions higher than 1 for any Hamiltonian with
non-singular low-energy interactions.$^6$  Therefore a principal part of the theoretical task
is to show that, in a model appropriate for Copper-Oxide metals, elimination of the high
energy degrees of freedom leads to a singular effective low energy Hamiltonian.  The
tremendous variety and number of experiments on Copper-Oxide metals severely constrain
the form of such a low energy Hamiltonian.

\subsection{Constraints From Experiments}

A schematic generic phase diagram is drawn
in Fig. (1) on the basis of the resistivity data.
Where measurements are available every other 
transport property
shows corresponding regions.  The insulating-antiferromagnetic phase near 1/2 filling and
the superconducting phase are shown in bold lines.  
The normal state is roughly divided into four regions with dashed lines representing
cross-overs from one characteristic temperature dependence in transport properties to
another.
Region 1 the 
non-Fermi-liquid phase has $\rho (T) \approx \rho_0 + \rho_1 T$ and similarly remarkable
``simple'' anomalies in all the
other transport properties.  In Bi 2201
resistivity measurements$^3$ are available from 10K to 800K and in
$La_{1.85} Sr_{.15} Cu O_4$ from 40K to 800K.$^7$  
The measured resistivity exponent in both is
$1.05 \pm 0.05$.  
We may safely assume that it is
1 with possible logarithmic corrections.  Such a behavior is observed only in a very narrow
region near $x_c$.  $\rho (T)$ begins to \it decrease \rm below the linear extrapolation as
temperature is decreased in region 3
and is consistent with an asymptotic $T^2$ dependence characteristic of a
Fermi-liquid.  On the low doping side,in region 2, there is a crossover to
resistivity increasing with decreasing temperature.  This regime may be termed insulating.
It is almost certainly related to singular effects of impurity scattering in a
non-Fermi-liquid (see Ref. (8) and Sec. (8) below).
Strictly speaking, one should draw a
third axis in Fig. (1) labelled disorder.  
If the zero-temperature intercept of the high temperature linear-resistivity is taken as
a measure of disorder one concludes that it generally increases in the available data as $x$ decreases.
The limited systematic data with independent
variation of disorder and x clearly shows
that impurities have a dramatic effect in the underdoped
regime while in the overdoped regime, region 3, their effect is conventional.$^9$

The size of the region 4 between region 1 and region 2 on the underdoped side depends upon disorder.
The crossover between region 1 and region 4 is marked by a decrease in
$C_v /T$ and $\chi$
with temperature with their ratio almost
independent of temperature.$^4$ In region 4 the
resistivity drops below the linear extrapolation from region 1.
This region has been termed the spin-gap region,
but this is a misnomer. Not only do magnetic fluctuations, but optical conductivity and Raman scattering intensity in all
measured polarizations decrease at low energies from their value in region 1. In region 2 this decline continues while the resisitivity begins to increase with decreasing temperature.

>From data for $YBa_2 Cu_4 O_8 (248)$
under pressure,$^{10}$ it appears that x is \it not \rm a unique parameter for
the crossovers in Fig. (1).  The stochiometric compound 248 has a $T_c \approx 80K$ and a
resistivity with a crossover from $\sim T$ to a higher $T$ dependence below about
200K, very
similar to the properties of $Y Ba_2 Cu_3 O_{6.7}$.  Under pressure, $T_c$ rises to 110K
at $P \simeq 100 kbar$.  Simultaneously the resistivity becomes linear down to $T_c$.
The low energy excitations also change under pressure.  In 248 the Raman spectra shows a low
energy decrease.
Under pressure as resistivity becomes linear the low energy spectrum is restored becoming
like that of the optimally doped $YBa_2 Cu_3 O_{6.9}$, i.e. frequency independent.$^{11}$

The schematic phase diagram of Fig. (1) suggests that the anomalous normal state (as
well as superconductivity) are controlled by fluctuations around the point
$x \approx x_c$ and $T \approx 0$.  This is consistent with the marginal
Fermi-liquid phenomenology$^{12}$ which suggests that 
the breakdown of Landau theory is due to scale invariant fluctuations
consistent with having a
quantum critical point,$^{13}$ QCP, (i.e. a singularity at $T = 0$) near the ideal composition.  The critical point itself is
inaccessible due to the superconductive instability. The nature of the symmetry breaking at the critical point was not specified.

The MFL has a single particle
self-energy of the form,
\begin{equation}
\Sigma (\omega , q) = \lambda 
\left [ \omega \ell n
\frac{\omega_c}{x} + i x sgn \omega \right ]
\end{equation}
where $x = max |\omega|$ for $|\omega| >> T$ and $= \pi T$ for $T >> |\omega|$,
$\lambda$ is a coupling constant and $\omega_c$ is a
cut-off energy.  The quasi-particle renormalization amplitude
\begin{equation}
z (\omega) = \left ( 1 -
\frac{\partial Re \Sigma}{\partial \omega} \right )^{-1} =
\left (1 + \lambda \ell n 
\frac{\omega_c}{x} \right )^{-1}
\end{equation}
then vanishes logarithmically as $( \omega , T ) \rightarrow 0$.

A microscopic theory should specify the nature of the critical
point and the symmetry on either side of it.
It should also answer the question:
If there is a critical point at $x=x_c$, $T=0$,
what about its continuation in the ($x$-$T$) plane?
Shouldn't there be evidence of nonanalytic properties on a line
in the ($x$-$T$) plane?
Experimentally, there is indeed a cross-over in the properties in the
($x$-$T$) plane from Region 1 to Region 4 of fig. (1).
But why a cross-over rather than a transition? Or is it that the properties 
studied like transport and specific heat are often weakly sensitive to
a transition?

Equation (1) gives only the single particle scattering rate $\tau_{sp}^{-1}$.  
This was used to understand$^{12,14}$
the observed tunneling conductance $G (V) \sim | V |$ for
$T \rightarrow 0$ and to predict the lineshapes in
single particle spectra.$^{15}$
A crucial aspect of the properties of the
Cu-O metals is that the momentum transport scattering rates, measured in resistivity $\rho
(T)$ (as well as in optical conductivity$^{16}$ $\sigma ( \omega , T )$ 
and Raman cross-section$^{17}$ $S_R ( \omega ,
T )$) are also proportional to max $( | \omega | , T)$.  So is the energy scattering rate
$\tau_{en}^{-1}$ measured by thermal conductivity $\kappa (T)$.$^{18}$
The experimental result that at $(q , \omega) \rightarrow 0$
\begin{equation}
\tau_{sp}^{-1} \sim \tau_{mom}^{-1} \sim \tau_{en}^{-1} \sim T
\end{equation}
puts a strong constraint on theories.
The single particle scattering rate is required in general to be at least as singular as
the momentum scattering rate $\tau_{mom}^{-1}$.
The immediate conclusion is that the experiments require
$Im \sum ( \omega , k_F ) \sim \omega^{\alpha} , \alpha \leq 1$.
Angle resolved photoemission
should be used to put stricter bounds on the single particle self-energy than
have been done so far. 
But an easier way might be through the electronic spefic heat.
The electronic specific heat is directly related to the exact single 
particle Green's function.
For the marginal case, $\alpha =1$, i.e. Eqn. (1.1) gives 
$C_v \sim N(0) T(1+ \lambda ln \omega_c / T)$.
A more singular self energy, 
$\alpha < 1$ gives
$C_v  \sim N(0)T^{\alpha}$.
In the experiments the electronic part of $C_v$ is obtained 
only by subtracting the estimated phonon heat capacity and at $x \approx x_c$ is 
reported to be consistent with $\sim T$. While logarithmic corrections to it 
cannot be ruled out, substantial singular departures are ruled out.

The proportionality of the single particle
and the transport scattering rates occur if the
fluctuations leading to (1.1) are essentially
momentum independent as suggested by the MFL phenemenology. (In that case there are no vertex corrections in the calculation of the conductivity.
See further discussion in Sec. (7)). 
But this poses the serious dilemna that on the one hand we wish to be near a critical
point, on the other that we need (nearly) momentum independent fluctuations.

The trivial way to get (1.3) is if the experiments are in a temperature range
$T \gtrsim \omega^*$ where $\omega^*$ is the characteristic frequency of some fluctuations
which scatter the Fermions.  Then the density of such fluctuations is
$\sim T$ giving (1.3).
This is ruled out by the specific heat (and magnetic
susceptibility experiments) experiments.$^4$  If such
fluctuations are of physical quantities like spin or density fluctuations of Fermions, a
characteristic enhancement in $\gamma = C_v /T$ of $O ( E_F / \omega^* ) \sim O (10^2)$
must occur as in the heavy Fermion compounds.
Experimentally, the specific heat and magnetic susceptibility
at the ideal composition are consistent with Fermi-liquid behavior, $\sim T$ and constant
respectively, with no more than about a factor of 2 enhancements over non-interacting
electrons in the measured temperature range.  (The data however does allow for
logarithmic or small power law corrections.)  This as well as the fact that
$\sigma (\omega , T)$ and $S_R (\omega , T)$ behaves smoothly in the
range $\omega \lesssim T \lesssim \omega_c$ where $\omega_c \sim O ( \frac{1}{2} eV)$
suggests that there is no low-energy
scale near optimum doping and that the upper cut off
frequency of the fluctuations is very high, $O (\frac{1}{2} eV)$.

It is hard to imagine that the fluctuations due to the antiferromagnetic $ T = 0$ critical
point at $x \approx .02$ can have much to do$^{19}$
with phenomena at $x \approx x_c$,
with $x_c \gtrsim 0.15$. Temperature independent magnetic 
correlation lengths of
about $2 \AA$ are observed at $x \approx x_c$ in $YBa_2 Cu_3 O_{6.93}$$^{20}$.
Temperature-dependent lengths of $\sim 20 \AA$ at $x \approx x_c$ in $La_{1.85} Sr_{.15} Cu
O_4$$^{21}$ but
with less than $\sim 10\%$ of the total frequency integrated spectral weight in the q-dependent part.
The normal state anomalies are identical in both compounds.  Similarly, there is no evidence
of any universal phase-separation fluctuations or charge density fluctuations$^{22,23}$
in different compounds at $x \approx x_c $.  It
would appear that if a critical point is
responsible for the unusual metallic state, it must be associated with some quite 
unusual order parameter which is hard to detect.

Equations (1.1) and (1.3) cannot be used to understand the observed anomalies in NMR,$^{24}$ 
Hall effect,$^{25}$ and magneto-resistance.$^{26}$
The anomalies in the Hall conductivity, $\sigma_{x y}$ and magnetoresistance are quite
revealing.  In $Y Ba_2 Cu_3 O_{6.9}$, where bandstructure calculations$^{27}$
give a very small usual
Hall conductivity due to particle-hole symmetry, the Hall angle $\Theta_H = \sigma_{xy} /
\sigma_{x x}$, with magnetic field perpendicular to the plane, varies$^{25}$ 
approximately as
$T^{-2}$ between 100K and 300K  In the same range the normalized magnetoresistance
$\Delta \rho (H) / \rho $ varies$^{26}$ roughly
as $T^{-4}$ with $\Theta_H ^2 / \Delta \rho / \rho \approx 0
(1)$.
The Hall effect data$^{25b}$ in $La_{2-x}Sr_xCuO_4$ however
appears more complicated where a saturation in the anomalous contribution
occurs at temperatures below about 60 K. In view of this, it is not clear 
at the moment whether the Hall-effect anomaly is aleading low temperature
singularity or an intermediate to high-temperature phenomena. It is 
worth noting that
a singularity in the Hall angle $\sim T^{-2}$, equivalently a Hall number
diverging as $T^{-1}$, implies a spontaneous Hall  effect in the limit
$T \rightarrow 0$, i.e. a Hall voltage in the absence of a magnetic field.    
Kotliar, et al.$^{28}$ have found this
behavior in a solution of the Boltzmann equation
which besides a scattering rate $\tau_{tr} ^{-1} \sim T$, contains a phenomenological skew 
scattering rate proportional to the applied magnetic field which is
$\sim T^{-1}$.  

Perhaps the most astonishing of the normal state anomalies are the nuclear relaxation rates
$^{Cu} T_1 ^{-1}$ and $^0 T_1 ^{-1}$ of copper and oxygen nuclei respectively.$^{24}$  While
$( ^{Cu} T_1 T )^{-1}$ appears to diverge as temperature is decreased suggesting singular \it
local \rm magnetic fluctuations at the O nuclei, 
$( ^0 T_1 T)^{-1}$ is a constant, which is the conventional behavior.
Furthermore, $(^0T_1 T K )^{-1}$, where K is the measured Knight 
shift at oxygen, is within 20\%
a constant, irrespective of the compound studied or the density x in any given compound in
the metallic range.$^{29}$  Other experiments show that copper and oxygen orbitals are well
hybridized.  Nevertheless, it appears the local magnetic fluctuations 
at copper and oxygen sites are quite different. The conventional behavior 
on Oxygen, independent of x, rules out the scenario that the anomalies on
Copper are
due to the peaking of anti-ferromagnetic fluctuations.

It is axiomatic that the fluctuations responsible for the anomalous metallic 
state also are
responsible for the instability to the superconducting state.
The anomalous fluctuations
develop a gap in the superconducting state as predicted$^{12a,30}$ 
and observed in a wide variety of
experiments on quasiparticle relaxation rate deduced through transport 
experiments$^{31}$ and in
angle resolved photoemission experiments.$^{32}$  
The symmetry of the superconducting state appears
to be consistent with ``D-wave'' (if the lattice is assumed tetragonal).$^{33}$
This issue is not completely settled yet.$^{34}$
Moreover the electron-doped material $Nd_{2-x}Ce_xCuO_4$
appears to be an ``S-wave''
superconductor.$^{35}$

To summarize, existing experiments require an {\it internally 
consistent} microscopic theory to: 
\\
(i) reproduce the phase-diagram of Fig. (1) with a non-fermi-liquid metallic phase near
the composition for the highest $T_c$ with cross-overs to fermi-liquid on the high doping  side and
with a strong tendency to insulating behavior due to disorder in the underdoped regime.  
The underdoped regime shows loss at low energies of both
particle-hole excitations (in spin as well as charge channels)
and of single particle excitations.
\\
(ii) have 
equilibrium properties like specific heat and magnetic
susceptibility
near the ideal composition
{\it consistent} in the measured range of $T$ with characteristic
Fermi-liquid behavior to within small corrections and in magnitude be within
factors of $O (2)$ of those for non-interacting electrons.
\\
(iii) have long wave-length transport relaxation rates used to interpret
electrical conductivity and thermal
conductivity at the ideal composition that satisfy Eq. (3).
At $x=x_c$ the fluctuations leading to the anomalous transport
should have no scale other than a cutoff of the
$0(1/2~eV)$.
\\
(iv) The fluctuations should have a symmetry such that they produce singular local magnetic fluctuations at copper nuclei
to give the observed anomalies in the copper nuclei relaxation rate, but no singular local
magnetic fluctuations at the oxygen nuclei,
\\
(v) The fluctuations should be capable of producing a  pairing instability of D-wave symmetry.  

There are of course many other special properties discovered in a
subject in which $O(5x10^4)$ papers have been published.
But I regard the requirements listed above as the most basic or a least the
irreducible minimum. The theory developed in this paper attempts to meet 
these requirements and suggest a few crucial experiments.

\subsection{Choice of a Model}

The choice of a model with which to do microscopic theory should be
influenced by the fact that copper-oxide metals are unique.  None of the thousands of
transition metal compounds studied share their properties.
The point of view taken here
and elsewhere$^{36}$ is that the unique properties of Cu-O metals arise from their unique
chemistry in which ionic interactions play a crucial \it dynamical \rm role.
This point has been
extensively discussed$^{37}$ and will only be briefly repeated here.
The divalent transition
metal oxides at 1/2-filling can be
put on
the diagram$^{38}$ in 
Fig. (2) in which one of the axes
is the normalized local repulsion energy $U$ on the transition metal
\begin{equation}
U/W = \left [ E (TM)^{3+} + E(TM)^{1+} - 2E (TM)^{2+} \right ] / W
\end{equation}
where W is the bandwidth.  The other axis is the \it ionic energy \rm
\begin{equation}
E_x / W = \left [ E (TM^{1+} O^{1-} ) - E (TM^{2+} O^{2-} ) \right ] / W
\end{equation}
$U$ is the energy to convert two transition metal ions with formal charge state
$2^+$ to one with formal charge state $1^+$ and the other to $3^+$, while $E_x$ is the
energy to transfer charge from the ground state configuration of a transition metal ion
with charge $2^+$ and a nearest neighbor oxygen ion with $2^-$ to transition metal ion
with charge $1^+$ and oxygen ion with charge $1^-$.
Screening and dipole corrections,
etc. in the solid are included in the definitions of $U$ and $E_x$.

As one moves from the left to the right of the periodic table, the ionization energy of the
TM falls thereby decreasing $E_x / W$ relative to U/W with corresponding movement on Fig.
(2).  In the insulating state of Cu-O, $E_x$ is only about 1eV.  It is a charge-transfer
insulator with the lowest energy one particle spectra
primarily on copper, $Cu^{2 +} \rightarrow Cu^+$ while the one-hole spectra is primarily
in oxygen, $O^{2-} \rightarrow O^{1-}$; see Figs. (3a and 3b) where the contrast to transition
metal oxides towards the left of the periodic table is also shown.  
In the metallic state, obtained by doping, there
are charge fluctuations on copper and on oxygen of similar magnitude.  $E_x \approx 1eV$ is made up of from
two sets of energies, the atomic level energies and the Madelung or ionic energies each of
which is $O (10 eV)$.  Indeed all transition metal oxides owe 
their structural stability to the ionic
energies.  But in the metallic state these energies have little dynamical role to play in
most TM-oxides because there are hardly any fluctuations on oxygen, i.e. although the ionic
fluctuation energy term in the Hamiltonian
\begin{equation}
\aSU_{i<j} \: V_{ij} \: n_i \: n_j
\end{equation}
has $\aSU_{i<j} V_{ij} \sim O (10 eV)$, the fluctuation $\delta n_{0j}$ 
on the oxygen ions require a large
energy and are insignificant.  An effective low energy Hamiltonian of the Hubbard form is
then adequate.  This is not true in the metallic state of Cu-O where
$< \delta n_0 > / < n_0 > \sim 0 (1).$  

One of the aims of this paper is to 
show that finite-range interactions, if sufficiently strong, lead to qualitatively
new features in the phase diagram of the model.  
The one-dimensional version of the model has been extensively investigated by numerical
methods.$^{39}$
Bosonization methods give incorrect results for the model
for $V's \gtrsim 0 ( E_x )$
whereas as is well known the one-dimensional Hubbard model can be bosonized for any
value of $U$.  Therein lies a clue to understanding how in one and higher dimension the
low-energy properties of the model are quite different for small $V$ and for large $V$.
Bosonization is inherently a weak-coupling method; it works in the (1D) Hubbard model
because the properties at weak-coupling (small $U/t$) are {\it similar} to those at
large-coupling.  The numerical results show a change in properties as $V$ is varied; for
$V \gtrsim 0 ( E_x)$ they show the charge transfer instability described below,
and growing superconducting correlation length as temperature is decreased, whereas
Bosonization methods find $V$'s to be irrelevant.  (1D-models, do not have some
essential features, discussed below, of the model in higher dimensions.)

\subsection{Preview of the Properties of the Model}

In Sec. (2) we discuss that it is enough to
consider a two band model representing Cu
and O bonding -b and anti-bonding -a
bands (but with O-O hopping included) as illustrated in Fig. (4).
In the hole representation, the chemical potential $\mu$
is in the lower band as shown.
For non-interacting electrons $\mu$ would be in the middle of band a at 1/2
filling as in $La_2 Cu O_4$ or $YBa_2 Cu_3 O_6$.  For hole doping, as for most Cu-O
compounds, $\mu$ rises with doping $x$, for example in $La_{2-x} Sr_x Cu O_4$ or $YBa_2
Cu_3 O_{6+x}$.
We will find it important to consider
the most general form of two-body interactions
allowed by symmetry
in the space of these two bands. When the
strength of the interactions is on the same scale as the overall electronic
bandwidth every term has a crucial role to play.

If $\mu$ were in the gap between bands a and b, the model is identical to the Excitonic
Insulator problem,$^{40}$ which has been extensively discussed.  We investigate here the model
with $\mu$ in one of the bands.  This changes the problem substantively.

It is well known that interactions completely alter the one-electron picture at
1/2-filling; a gap develops around the chemical potential and the relative amount of Cu
and O character of the occupied and un-occupied states is drastically altered.
$\mu$ stays in the band
significantly away from 1/2-filling
but the relative Cu-O
character of the occupied and unoccupied states is again expected
to be quite different from the one-electron
picture.  If no change in the lattice symmetry occurs due to the interactions, this is
formally describable by a $q = 0$ instability of the one-electron band structure to a
state in which
\begin{equation}
T_x \sim Re \aSU_{k}
{\cal F}_x (k)
\left < a_{k \sigma}^+ b_{k \sigma} \right > \neq 0
\end{equation}
Here ${\cal F}_x (k)$ is a form factor expressing the relative Cu and O character of states
in bands a and b.  We will see that $T_x$ is closely related to the relative average
charge in Cu and O orbitals.  The mean-field free-energy as a function of $T_x$ is shown
in Fig. (5a).  With $T_x \neq 0$, the orbitals must be
rehybridized leading to new bands $\alpha$ and $\beta$ of the
same general form as a and b, as shown in Fig. (4).  This is expected
to occur for strong interactions at very high temperatures for all $x$ of interest just
as it does at $x = 0$.  For the most general interactions, such a transition is of first
order, as for a free-energy of the form shown in Fig. (5a).
Previous investigations$^{41,42}$ of
the model had focussed on this charge transfer instability.

We show that the model also has an interesting second-order transition in
the Ising class to a state in which
\begin{equation}
T_y \sim Im \aSU_{k \sigma} {\cal F}_y(k) 
\left < a_{k \sigma}^+ b_{k \sigma} \right >
\neq 0  .
\end{equation}
The mean-field free-energy, for a fixed $T_x \neq 0$, as a function of $T_y$ for various
values of a parameter $p$ which is a function of $x$ and $T$ is shown in Fig. (5b).
This transition therefore
occurs on a line $T_c (x)$ in the $x - T$ plane.  We will be specially
interested in the properties in the vicinity of the point $x = x_c (0)$ where
$T_c = 0$ which we identify as the quantum critical point.$^{43}$

A finite $T_y$ provides an additional relative phase to the wavefunctions at Cu and the two O
sites in a unit cell.  We will see that the ground state with a finite $T_y$ corresponds
to a four-fold pattern of circulating current within a unit cell with all cells staying
equivalent. This is illustrated in Fig.(6).
We may call it the circulating current (CC) phase. Translational symmetry is preserved but time reversal-invariance and four fold rotational symmetry
are broken. The product of the time-reversal and the four-fold rotaion are
preserved.

In connection with the excitonic
insulator problem and the Hubbard model at 1/2-filling circulating current phases have
also been discussed, which also break translational symmetry,
and go by the names of orbital antiferromagnets$^{44}$
or staggered flux phases.$^{45}$
There are two important differences - 
(i) Our use of a Cu-O model with more than two atom per unit cell allows a $q =
0$ transition to a circulating current phase, so that lattice translation symmetries are preserved.  There is no change in the symmetry of the band-structure. 
(ii) We discuss such a phase in the metallic state.
This leads to a very special nature
of the collective fluctuations near the
QCP due to scattering of the
fluctuations by low-energy particle-hole excitations 
at the Fermi-surface.  The
problem of determining the spectrum of the fluctuations 
is the same as the absorption in a degenerate semi-conductor with finite mass electrons and holes
and in the limit that the interactions are much larger than the Fermi-energy. We show that the logarithm of the frequency of the fluctuations scales with their momentum - i.e the QCP has dynamical critical
exponent $z_d = \infty$. The fluctuations are thus independent of momentum
to a logarithmic accuracy. This extreme quantum limit is essential to understand the observed
behavior in long-wavelength -low frequency transport properties summarized by Eq.(1.3) and its finite frequency counterparts.

The model is stated more completely in Sec. (2).  One can study the model as
interactions are increased from zero or take a strong-coupling point of view and
consider corrections about an infinite value of the interactions.  Both approaches lead
to the same ``intermediate energy scale'' Hamiltonian whose analysis begins in Sec. (4).
Sec. (3) is devoted to the strong-coupling expansion.  
The low energy Hamiltonian is derived and analyzed in
Sec. (5) and (6) where it is shown that a systematic and controlled analysis is
possible.
The physical properties of the pure model are investigated in Sec. (7), and in
Sec. (8) a beginning is made to consider the effect of impurities in the properties of the
model.
Impurities are strongly relevant near a
$z_d= \infty$ transition.
For arbitrarily small concentration, they convert the line $T_c(x)$
to a cross-over.
We also show that for arbitrarily small concentration of
impurities in a nonFermi-liquid, the Fermi-surface
withers away at low temperatures.
The density of states at the chemical potential is zero and the 
resistivity is infinity as $T \rightarrow 0$ (unless
superconductivity intervenes).

Not including the effect of disorder the schematic deduced phase-diagram of the model is shown in fig. (7).
In region I, the properties are determined by quantum fluctuations and are that of a marginal Fermi-liquid
with a cross-over to a Fermi-liquid regime in region III.
In region II  $T_y \neq 0$ in the pure limit.
This phase should have fermi-liquid properties at low
temperatures in the pure limit but with different parameters from that of region III.
The transition between regions I and II turns into a crossover
at arbitrarily small concentration of impurities. At low temperatures impurities are expected to lead to a further crossover in region III to an insulating regime with
zero density of states at the chemical potential.
Antiferromagnetism phase near $1/2$-filling and
superconductivity phase near $x_c$ which are also properties of the model are not shown.

In Sec. (9), a beginning is made to study the pairing instability due to the
exchange of the circulating current fluctuations and a
propensity towards a ``D-wave'' superconducting instability is
indicated.

In Sec. (10) the shortcomings of the theoretical calculations
as well as the unexplained features of the experiments are
highlighted.
Experiments are suggested to test several features of the
theoretical proposal. the most important is to observe the current pattern
of fig.(6) by polarized neutron or x-ray scattering.

\setcounter{equation}{0}
\section{Model for Copper-Oxides}
The basic building block of the Cu-O compounds is the elongated
$CuO_6$ octahedra in which the planar-short bond oxygens are shared
at the corners to produce a layered, anisotropic three dimensional
structure.
The interlayer kinetic energy depends on the details of the structure.
In the least anisotropic compounds the inter-layer bandwidth
is $O(10^{-1})$ of
the intralayer bandwidth.
Since the properties of such compounds in the temperature
region of the normal state are the same as of those with
anisotropy ratio $O(10^{-4})$, a two-dimensional model is appropriate
for the essential physics.
The basis set for the minimum Hamiltonian is the
$d_{x^2-y^2}$ orbital on the Cu-ions and the
$p_x$ and $p_y$ orbitals on the O-ions -- see Fig. (8).
In this basis set the Hamiltonian is written as
\begin{equation}
H = H_0 + H_1 +H_2 .
\end{equation}
$H_0$ is the kinetic energy
\begin{equation}
\begin{array}{rl}
H_0 & = \; \aSU\_1 \frac{\Delta_0}{2}\;
(n_{di}-n_{pxi}-n_{pyi} ) -
\mu(n_{di} + n_{pxi} +n_{pyi} ) \\ 
& + \;
\aSU_{i, \sigma} t_{pd}\; d_{\sigma i}^+ \;
(p_{x, i + \frac{a}{2}, \sigma} - p_{x, i-a, \sigma} 
+ p_{y, i+ \frac{a}{2}, \sigma}
-p_{y, i,- \frac{a}{2}, \sigma}) \\
& + \; t_{pp}\; p_{x, i + a, \sigma}^+\; 
( p_{y, i + \frac{a}{2}, \sigma}
-p_{y, i - \frac{a}{2}, \sigma}) \\
& + \; h.c. \\
\end{array}
\end{equation}
Here ($i$) sums over the unit cells in the plane,
$d_{i\sigma}^+$ refers to the Cu-d-orbitals and 
$p_{(x,y),i \pm \frac{a}{2}, \sigma}$ to the
oxygen $p_x$ or $p_y$ orbitals, 
which are neighbors to the Cu in cell $i$ at a distance $a/2$.  
The relative signs in
the kinetic energy take into account the phases of the orbitals,
as shown in Fig. (8).

$H_1$ is the short-range part of the interactions
\begin{equation}
H_1 = \aSU_{i, \sigma}
U_d n_{di\sigma} n_{di \bar{\sigma}} +
U_p n_{pi\sigma} n_{pi \bar{\sigma}} .
\end{equation}

$H_2$ is the long-range part of the electron-electron
interactions and the exchange interaction
important only for nearest neighbors
\begin{equation}
H_2= \aSU_{i,j} V_{ij} n_in_j
+ \aSU_{(i,j)} V_x {\bf s}_i \cdot {\bf s}_j ,
\end{equation}
where $n_i=n_{di}$ or $n_{px, i}$, $n_{py, i}$ as appropriate.
One must include the long-range $V_{ij} \sim |R_i-R_j|^{-1}$ for
$|R_i-R_j| \rightarrow \infty$ to keep the long wavelength charge
oscillations at a finite plasma frequency.
Otherwise, only the nearest neighbor Cu-O and O-O
$V_{ij}$ need be considered.

Diagonalization of $H_0$ gives the model ``one-electron''
band-structure, which has ``Cu-O bonding'' and 
``anti-bonding bands'' and
a $O-O$ ``non-bonding band''. 
For $t_{pp} = 0$, the bonding band -b and the antibonding band -a
are decoupled from the non-bonding band.  $t_{pp} \neq 0$ is
important for the results of this paper, but I consider only the
bonding and the anti-bonding for simplicity and neglect the
non-bonding band.  The eigenvalues and eigen-vectors of the bands
kept must include the effects of $t_{pp}$.  It will be clear that
smaller interactions are needed in the three-band model than in
the simplified model for the important instabilities derived in
Sec. 4.

The dispersion of the bands a and b will be denoted by 
$\epsilon_a ( {\bf k} )$ and $\epsilon_b ( {\bf k} )$.  These are
sketched in Fig. (4).  Expressions for them, derived
perturbatively in $t_{pp} / t_{pd}$ are given in the Appendix A.
Their eigen-vectors are specified by the annihilation operators
for states in them in terms of annihilation operators for $d$ and
$p_{x,y}$ orbitals at lattice sites:
$$
a_{{\bf k} \sigma} = u_{ad} ({\bf k}) d_{{\bf k} \sigma}
+ u_{ax} ({\bf k}) \; sin \left( 
\frac{k_x a}{2} \right) p_{x {\bf k} \sigma} + u_{ay}
({\bf k}) sin \left( \frac{k_y a}{2} \right) p_{yk \sigma} 
\eqno{(2.5a)}
$$
$$
b_{{\bf k} \sigma} = u_{bd} ({\bf k}) d_{{\bf k} \sigma} +
u_{bx} ({\bf k})\; sin \left( \frac{k_xa}{2} \right) 
p_{x {\bf k}\sigma} + u_{by} ({\bf k}) sin
\left( \frac{k_ya}{2} \right) p_{y {\bf k} \sigma}
\eqno{(2.5b)}
$$
where
\[
\begin{array}{rl}
d_{k\sigma}^+ &= \frac{1}{\sqrt{N}} \aSU_i d_{i\sigma}^+ e^{ik\cdot R_i} \\
p_{x(y)k\sigma}^+ &= \frac{1}{\sqrt{N}} \aSU_i
p_{ix(y)\sigma}^+ e^{ik \cdot R_i}
\end{array}
\]
${\bf R}_i$ label the unit cells; the Cu atom in the unit cell $i$
is taken to be at ${\bf R}_i$.  Expressions for these coefficients,
also calculated perturbatively in $t_{pp} / t_{pd}$ are given in
Appendix A.

Bandstructure calculations give
$t_{dp}, 2t_{pp}, \Delta_0 \sim O (1 eV)$; 
various spectroscopic methods give
the local repulsion on Cu - $U \sim O(10 \; eV)$.
As good an estimate of the nearest neighbor Cu-O repulsion as any is
\setcounter{equation}{5}
\begin{equation}
V_{nn} \gtrsim  e^2 / [ \epsilon (1 eV) R_0 ] ,
\end{equation}
where $R_0$ is $1/2$ the unit cell size and $\epsilon (1 eV)$
is the measured long wavelength di-electric constant
at an energy of O(1 eV), which is
$\approx 4$.
This gives the nearest neighbor Cu-O interaction $\gtrsim 1.7 eV$.
So, multiplied by the number of neighbors, and considering
polarization corrections etc., the characteristic Madelung
energies
controlling Cu-O charge fluctuations is also of O(10~eV).
Calculations of Madelung energies$^{46}$  in the actual lattice support this estimate.

The interaction energies and the overall electronic bandwidth are
therefore of the same order.
We can directly
project the Hamiltonian (2.1) on to the basis set of the bonding
and anti-bonding k-space orbitals obtained by diagonalizing (2.2).
This results in the most general two-band Hamiltonian allowable by
symmetry.
Such a Hamiltonian (projected to interactions in the spin-singlet channel,
which alone is important) is given in Eq. (4.1).
The reader may at this point skip directly to Eq. (4.1) and the subsequent
analysis.
Although this approach is quite consistent for the metallic state, it is hard
to derive the insulating phase near $1/2$-filling from such a basis
or to see that although the low energy physics at $1/2$-filling
of the general model is identical to the Hubbard model, it
may not be so in the metallic state.
A basis of local real-space orbitals constructed in the strong-coupling
limit to suppress some of the charge fluctuations is more convenient.
The next section is devoted to deriving the projections
of the Hamiltonian (2.1) to such a basis.
The subsequent analysis, which is a simple generalization of slave
Boson methods$^{47}$ yields the Hamiltonian (4.1) as well.
The results of this paper are qualitatively similar starting from either end
in the ratio of interaction energy to the one-particle bandwidth.
\section{The Strong-Coupling Limit}
\setcounter{equation}{0}
It is convenient
to rewrite the Hamiltonian as an intracell part and an
intercell part
to calculate in the strong-coupling limit.
We define a linear combination $D_{i\sigma}^+$ of operators on
oxygen orbitals in a cell $i$ which hybridize with the
$d_{x^2-y^2}$ orbital in the same cell:
\begin{equation}
D_{i\sigma}^+ = \frac{1}{\sqrt{4}}
(p_{i+a_x} - p_{i-a_x}+
p_{i+a_y}-
p_{i-a_y} ) .
\end{equation}

$D_{i\sigma}^+$ creates
an orbital which also transforms as a $d_{x^2-y^2}$
orbital about the center of the cell $i$.
In the geometry of the Cu-O lattice the orbitals created
by $D_{i\sigma}^+$ are not orthogonal for near-neighbor $i$.
Wannier orbitals
$\omega_{i\sigma} ( {\bf r} )$
can be defined which are orthogonal for different $i$.
Such orbitals can be expressed as a linear combination of
the orbitals created by $D_{i\sigma}^+$.

$H_0$ can be re-expressed as
\begin{equation}
H_0 = H_{0 , \; cell} + H_{0, \; inter-cell} ,
\end{equation}
where because there is no hybridization at the zone center
\begin{equation}
H_{0, \; cell} = \aSU_i \Delta_0 (n_{di}-n_{D i}) .
\end{equation}
We write
the kinetic energy completely generally as
\begin{equation}
H_{0, \; inter-cell} =
\aSU_{(ij),\sigma ,\mu \nu}
t_{i\mu , j \nu}
(\mu_{i\sigma}^+ \nu_{j\sigma} + h.c. )
\end{equation}
so that it reproduces the one-electron
bonding and anti-bonding bands.
Here ($\mu , \nu$) sum over $d$ or $D$.
As already discussed, use of non-orthogonal orbitals makes
$t_{ij}$ non-zero over a range larger than nearest
neighbors.
This conflict between the necessity of using non-orthogonal orbitals
to handle strong local, but not just on-site, interactions and
Bloch waves for a periodic lattice appears unavoidable.
We need not dwell on this because the final projected Hamiltonian (4.1)
for further analysis depends only on symmetry.

The interaction terms $H_1$, and $H_2$, remain
of the same general form in terms of $D_{i\sigma}$'s as in
Eq. (2.3) and Eq. (2.4) with redefinition of coefficients.
We will simply regard that the interaction terms
have been re-written in terms of D's, without
changing the notation for the new coefficients.
\subsection{States in the Strong-coupling Limit}
The low-energy Hamiltonian for this model will now be derived.
Since some of the important interactions are intracellular we specify a
basis set in terms of the states of a cell by cutting off the
kinetic-energy connection between cells and the long-range Coulomb
interactions.

Consider an average occupation of ($1+x$) holes
per unit cell as required by $(1+x)$ negative charges per unit
cell assumed uniformly distributed by imposing a chemical potential $\mu$.
The minimum {\it low energy} basis must then include states with
one-hole and with two-holes per unit cell.

We define the {\it zero-hole state}
$\phi_{0i}^+$
$|0>$ as the closed-shell (spin zero)
configuration in which the charge state of all oxygen ions is
$0^{- -}$ and of all copper ions in
$Cu^+$. \\
{\it One-hole States:}
These are of two kinds:
\begin{itemize}
\item[(i)]
$d_{1\sigma i}^+ |0>$;
a hole in the Cu-$d_{x^2-y^2}$ orbital with energy
$\ \Delta_0 - \mu$.
Chemically, this is the spin $1/2$ state $Cu^{++} O^{- -}$ which
is the nominal ground state configuration of the insulator.
\item[(ii)]
$d_{2\sigma i}^+ |0>$:
a hole in the orbital created by $D_{i\sigma}^+$,
i.e. the oxygen-d-orbital, with energy $- \Delta_0 - \mu$.
(If such an orbital were localized on one-atom, the charge
configuration would be $Cu^+ O^-$.)
\end{itemize}
{\it Lowest Energy Two-hole States $\phi_i^+ |0>$.}
The lowest energy two-hole state is the spin-singlet state with
one-hole in the Cu d-orbital and the other in the oxygen
d-orbitals.
The energy of this state is $E_{\phi} \equiv V-2 \mu$.
(If the latter were confined to one-atom, the charge configuration
would be $Cu^+O^-$.)
$\phi_i^+$ should be thought of as a hard-core Boson operator. \\
{\it Neglected States:}
These are
\begin{itemize}
\item[(i)]
Two-holes in the bonding combination of oxygen orbitals
with energy $-2 \Delta_0 -2 \mu +U_p$.

No essential physical difference arises if we include
these two-hole states also in the low-energy sector.
\item[(ii)]
Triplet state with one-hole on Cu-d and the other on the oxygens.
This is above $\phi_i^+ |0>$ by the exchange-energy,
$V_x$ which is $O(V_{ij})$
for $(ij)$ nearest neighbors.
\item[(iii)]
Two holes on the Cu-d orbitals with energy
$2 \Delta_0 -2 \mu +U_d$.
\item[(iv)]
Three or higher number of holes per unit cell.
\item[(v)]
The zero hole state
$\phi_{0i}^+ |0>$
with energy 0.
\end{itemize}

At low energies, the allowed cells in a cell $i$ must fulfill
the completeness relation or constraint
\begin{equation}
\psi_i^+ \psi_i +
\phi_i^+ \phi_i =1
\end{equation}
where $\psi_i \equiv (d_{1\uparrow} \;  d_{1\downarrow} \;
d_{2\uparrow} \; d_{2\downarrow} )_i$.
It is convenient to introduce Pauli matrices
$\mbox{\boldmath $\sigma$} $
and $\mbox{\boldmath $\tau$}$ to specify respectively the spin
($\uparrow \; \downarrow$) degree of freedom and the orbital (1 2) degree
of freedom in the one-hole sector of the problem.

To derive a low-energy Hamiltonian, we must project (2.1) to states
which fulfill the constraint (3.5).
To this end,
the bare operators $d_{i\sigma}$ and $D_{i\sigma}$ are expressed in
terms of the constrained operators through the identities
\begin{eqnarray}
d_{i\sigma}^+ &= \frac{1}{\sqrt{2}}\; \phi_i^+ d_{2i-\sigma} sgn \sigma \\
D_{i\sigma}^+ &= \frac{1}{\sqrt{2}} \; \phi_i^+ d_{1i-\sigma} sgn \sigma .
\end{eqnarray}
The {\it intra-cell} terms are transformed by noting that
\begin{eqnarray}
n_{di\sigma} &= n_{1i\sigma} + n_{\phi i /2} \\
n_{D_{i\sigma}} &= n_{2i\sigma} + n_{\phi i /2} ,
\end{eqnarray}
where
\begin{equation}
n_{1\sigma} = d_{1\sigma}^+ \;  d_{1\sigma} \; etc., \; and \;  \;
n_{\phi_i} = \phi_i^+ \phi_i .
\end{equation}

\subsection{Hamiltonian in the Strong-coupling Limit}
In terms of allowed states in the cell
\begin{equation}
V_0 n_{di} n_{D i} = V_0 n_{\phi i} ,
\end{equation}
and for $i \neq j$,
\begin{equation}
\begin{array}{rl}
V_{ij} n_{di} n_{Dj} &= V_{ij} (n_{1i} n_{2j} +
n_{\phi i} n_{\phi j} \\ \nonumber
                     &+ n_{1i} n_{\phi j} + n_{\phi i} n_{2j} ) .
\end{array}
\end{equation}
The term $\aSU_j V_{ij} n_{1i} n_{2j}$ is assumed already included in the
definition of the difference $\Delta_0$ of the one hole states.
Summing over $i$ and $j$ and using (3.5) the last two terms cancel the
second term and renormalize $E_{\phi}$.
In general, the interactions of the $d_1$ state and the
$d_2$ state with the neighbors are different and symmetry allows terms
of the form
\begin{equation}
\aSU_{(i,j)} \bar{V}_{ij} (n_{1i} - n_{2i}) \; n_{\phi j}
\end{equation}
These renormalize $\Delta_0$ downward proportionally to the density
of two-hole states as found in Hartree-Fock and
other previous calculations.$^{41,42}$
Equation (3.3) is simply,
\begin{equation}
H_{0,cell} = \Delta_0
\aSU_i (n_{1i} - n_{2i})
\end{equation}

Consider now the inter-cell part of the kinetic energy.
Starting from a configuration obeying the constraint, the inter-cell
kinetic energy leads to configurations which preserve the constraint
as well as those that do not.
Consider first the former.
These are necessarily processes
which alter the two hole occupation in cell $i$ and
one hole occupation in cell $j$ to one hole in $i$ and
two hole in $j$ or vice-versa, for example
\begin{equation}
\phi_i^+ d_{1j\sigma}^+ |0>  \rightarrow d_{2i\sigma}^+ \phi_j^+ |0> .
\end{equation}
Therefore, Eq. (3.4) projected to the lower energy states gives
\begin{equation}
H_{0,intercell} = \aSU_{(ij), \sigma , \mu \nu} t_{i \mu ,j \nu} 
( \phi_i^+ d_{\mu \sigma i}
d_{\nu \sigma j}^+ \phi_j + h.c. )
\end{equation}
The kinetic energy also operates on the one-hole states of
neighboring cells $i$ and $j$,
creating disallowed states $\phi_{0i}^+$ and the
disallowed two-hole states on j.
Eliminating such a kinetic energy term
by a {\it canonical transformation}
leads to an effective low energy
interaction in the space of the allowed one hole states.
This process is similar to that by which a Heisenberg exchange
Hamiltonian is generated from the Hubbard Hamiltonian.$^{48}$
The new feature here is that the one-hole sector has a
$\mbox{\boldmath $\tau$} $ degree of freedom as well as a
$\mbox{\boldmath $\sigma$} $ degree of freedom.
So there is an exchange in $\mbox{\boldmath $\tau$ } $ space as well
as $\mbox{\boldmath $\sigma$ }$ space.
The one-band Hubbard model produces an isotropic exchange Hamiltonian
because the full Hamiltonian is invariant to spin rotations.
This is again true in $\mbox{\boldmath $\sigma$}$ space here, but not in
$\mbox{\boldmath $\tau$}$ space.
The eigenvectors of the pseudo-spin $\mbox{\boldmath $\tau$}$
have in general different local energies,
$\Delta \neq 0$ in (3.14), and different transfer integrals.
Also, there are very many different intermediate states
with no obvious rotational invariance.
Here we derive the form of the effective interaction
Hamiltonian from completely general considerations.
An explicit derivation with calculation of the coefficients
is given in the Appendix.

The intracell kinetic energy terms in (3.4) which
connect the allowed one hole states to the  disallowed
states may be written in terms of products of operators
\begin{equation}
\phi_{0i}^+ d_{\alpha \sigma i} , \; \;
\phi_{2\mu j}^+ d_{\alpha \sigma j} , \; \alpha = 1,2
\end{equation}
where $\phi_{0i}^+$ create the zero-hole state
and $\phi_{2\mu i}^+$ creates one of the disallowed two-hole
states labelled by $\mu$.
The canonical transformation consists in eliminating the
intermediate states $\phi_{0i}^+ \phi_{2\mu j}^+$
(where now the allowed two-hole state $\phi_j^+$ is
included in $\mu$).
The most general {\it pair-wise} effective Hamiltonian
is the sum over products of two kinetic energy
operators with appropriate energy denominator.
It has the general form
\begin{equation}
H_{in t}= \aSU_{ij} J_{ij} \aSU_{\sigma , \tau}
( x_{\tau} \psi_{i\sigma \tau}^+ \psi_{j\sigma \tau} )
\aSU_{\sigma^{\prime} , \tau^{\prime}}
(x_{\tau^{\prime}}
\psi_{j\sigma^{\prime} \tau^{\prime}}^+
\psi_{i\sigma^{\prime} \tau^{\prime}})
\end{equation}
where ($\sigma = \uparrow , \downarrow$) and
$\tau= (d_1,d_2)$.
In (2.23) $J_{ij} x_{\tau} x_{\tau^{\prime}}$ is the sum over
intermediate high energy (disallowed) states of matrix
elements to such states divided by the corresponding
energy denominators.

Equation (3.18) can be rewritten as
\begin{equation}
= - \aSU_{ij} J_{ij} ( 1/4- \mbox{\boldmath $\sigma$}_i \cdot
\mbox{\boldmath $\sigma$}_j )
({\bf  A} / 4- \mbox{\boldmath $\tau$}_i
{\bf A} \mbox{\boldmath $\tau$}_j ) .
\end{equation}
The Hamiltonian is isotropic in
$\mbox{\boldmath $\sigma$}$ space and ${\bf A}$ expresses
the anisotropy in $\mbox{\boldmath $\tau$}$ space:
\begin{equation}
A_{\tau \tau^{\prime}} =
x_{\tau} x_{\tau^{\prime}}
\end{equation}
In (3.19)
\begin{eqnarray}
\tau_{zi} &=& (d_{1}^+ d_{1} -
d_{2}^+ d_{2} )_i , \\
\tau_{xi} &=& (d_{1}^+ d_{2} +
d_{2}^+ d_{1} )_i \\
\tau_{yi} &=& i ( d_{1}^+ d_{2} -
d_{2}^+ d_{1} )_i .
\end{eqnarray}
With axes defined as in (3.21)-(3.23) the most general
form of ${\bf A}$ is such as to generate:
\begin{eqnarray}
H_{i nt} &=& H_{xy} + H_{anis} , \\
H_{xy} = [ J_{zz} \tau_z^i \tau_z^j &+& J_{\perp}
( \tau_+^i \tau_-^j + h.c.) ]
(1/4- \mbox{\boldmath $\sigma$}_i \cdot\mbox{\boldmath $ \sigma$}_j), \\
H_{anis} &=& (J_{zx} (\tau_x^i \tau_x^j +
\tau_x^i \tau_z^j )
+ J_{\perp}^{\prime} (\tau_+^i \tau_+^j +
\tau_-^i \tau_-^j ) )
(1/4 - \mbox{\boldmath $\sigma$}_i \cdot\mbox{\boldmath $ \sigma$}_j) .
\end{eqnarray}
The only conceivable terms missing in (3.24-3.26) are those linear in
$\tau_y$.
Note that
$\tau_{yi}  =i$
($D_i^+ d_i- d_i^+ D_i)$.
Therefore
(remembering that $D_i$ and $d_i$ refer to wavefunctions at different
points in the unit cell $i$)
$\tau_{yi}$ represents a current distribution
within the unit cell $i$.
Terms linear in $\tau_{yi}$
can not be generated from a  time-reversal
invariant Hamiltonian.
Another way of seeing this is that if one has two bands
as in fig. (4) whose states are created by linear combination
of operators $a_i^+$ and $b_i^+$ respectively, the
most general two-body interactions (with operators on sites i and j) are
(ignoring spin)

\[
\begin{array}{rrr}
a_i^+ a_j^+ a_ja_i , & b_i^+b_j^+b_jb_i ,\; &a_i^+b_j^+b_ja_i  \\
a_i^+ b_j^+ a_jb_i & a_i^+a_j^+b_jb_i  \\
a_i^+a_j^+a_jb_i , & b_i^+b_j^+b_ja_i 
\end{array}
\]
plus Hermitian conjugates of these.
The terms in the first line can be rewritten in terms of
$\tau_z^i \tau_z^j$, in the second line in terms of
$( \tau _+^i \tau_-^j + h.c.)$ and
$( \tau_+^i \tau_+^j + h.c.)$ and those in the third
line in terms of ($\tau_{zi} \tau_{xj} + h.c.)$ just as in (3.24-3.26).

At $1/2$-filling, $x=0$, the state $d_{2 \sigma i}^+ |0>$ is not allowed
in the low-energy subspace.
Then $\psi_i \equiv (d_{1\uparrow} d_{1\downarrow})$ only, and the
low-energy Hamiltonian is obtained by dropping the
$\mbox{\boldmath $\tau$}$ dependence in (2.24).
The familiar Heisenberg Hamiltonian is then obtained.
If one drops the $\mbox{\boldmath $\tau$}$ variable in the
metallic state as well, the familiar $t-J$ Hamiltonian derivable
from the Hubbard model in the strong-coupling limit is obtained.
As discussed in the introduction, this is not justifiable for the
parameters of the Cu-O problem.
The $\mbox{\boldmath $\tau$}$ degrees of freedom make the
problem richer and afford the possibility of new physics
pursued in this paper.

At this point it is useful to collect all the terms of the effective
Hamiltonian
\setcounter{equation}{26}
\begin{equation}
\begin{array}{rl}
H &= \aSU_i (\Delta_0 + \aSU_j
V_{ij} n_{\phi j})
(n_{1i}-n_{2i})
- \aSU_i \lambda_i (n_{1i}+n_{2i}+n_{\phi i} -1)
- \mu \aSU_i (n_{1i}+n_{2i}+n_{\phi i} ) \\
  &+ H_{0, intercell}
+ \aSU_i E_{\phi} n_{\phi i} 
+ H_{i nt} .
\end{array}
\end{equation}
$\lambda_i$ enforces the constraint and $\mu$ is introduced to fix
the hole density at $(1+x)$.
$H_{i nt}$ is given by (3.24)-(3.26).
\subsection{Mean Field for the Slave Bosons}
We look for uniform mean-field solutions
\begin{eqnarray}
\lambda_i &= \langle  \lambda_i \rangle = \lambda \\
\phi_i &= \langle \phi_i \rangle = \phi .
\end{eqnarray}
We also look for spin-singlet solutions in
the bonds ($i-j$) favored by the kinetic energy term in (3.27) and
the ``RVB'' decomposition.$^{49}$
\begin{equation}
\begin{array}{rl}
J &(\aSU_{\sigma , \tau} x_{\tau} \psi_{i\sigma \tau}^+
\psi_{j\sigma \tau} )
( \aSU_{\sigma^{\prime} \tau^{\prime}}
x_{\tau^{\prime}} \psi_{j\sigma^{\prime}\tau^{\prime}}
\psi_{i\sigma^{\prime}\tau^{\prime}}) \\
  &\approx \epsilon_{ij} (\aSU_{\sigma \tau}
x_{\tau} \psi_{i\sigma \tau}^+
\psi_{j\sigma \tau} ) +
\epsilon_{ij}^2 / 4J_{ij}
\end{array}
\end{equation}
where
\begin{equation}
\epsilon_{ij} = 2J_{ij}
\langle \aSU_{\sigma , \tau} x_{\tau}
\psi_{i\sigma \tau}^+
\psi_{j\sigma \tau} \rangle
\end{equation}
are mean-field amplitudes.

The other decomposition of the interaction term in
which mean-field amplitudes for $\mbox{\boldmath $\tau$}_i$ are introduced
is more important to us.
Note that given spin-singlets in the bonds ($i-j$), uniform spatial
solutions in $\tau$-space are favored by the interactions in Eqs. (3.24),
for $J_{\perp} > J_{\perp}^{\prime}$.

We now diagonalize the bilinear terms in $\tau$
space, i.e. the first five terms in Eq. (3.27) and transform to
$k$-space.
This introduces bands $a$ and $b$.
Let $a_{k\sigma}^+$, $b_{k\sigma}^+$ create particles
in these bands:
\begin{eqnarray}
a_{k\sigma} &=& u_k d_{1k\sigma} + v_k d_{2k\sigma} \\
b_{k\sigma} &=& -v_k d_{1k\sigma} + u_k d_{2k\sigma} \\
v_k /u_k &=& tan \left [ \frac{1}{2} tan^{-1}
\tilde{t}_k / \Delta \right ]\\
\tilde{t}_k &=& t_k \phi^2 \\
and \; \; \; \Delta &=& (\Delta + \bar{V} \phi^2 )
\end{eqnarray}
Here $t_k$ is the lattice momentum transform of $t_{ij}$.
The effective Hamiltonian projected to the bands $a$ and $b$ is
given by Eq. (4.1).
We denote the dispersion of the two bands due to this diagonalization also
by $\epsilon_{ka}$ and $\epsilon_{kb}$.
They have the symmetry properties of the bandstructure in the one-electron
approximation, i.e. Eqn. (2.5).
\section{Analysis of the Two Band Hamiltonian}
\setcounter{equation}{0}
In the strong-coupling limit the {\it intermediate} energy scale
Hamiltonian obtained from (3.27) after (3.28) and (3.29) is
\begin{equation}
\begin{array}{rl}
H = &- (\lambda+ \mu -1) \aSU_{k\sigma}
(a_{k\sigma}^+ a_{k\sigma} + b_{k\sigma}^+ b_{k\sigma} )
-  (\lambda + \mu  - E_{\phi} ) n_{\phi} \nonumber \\
    &+ \aSU_{k,\sigma} \epsilon_{ka\sigma} a_{k\sigma}^+
a_{k\sigma} + \epsilon_{kb\sigma} b_{k\sigma}^+
b_{k\sigma}
+ H_{i nt} .
\end{array}
\end{equation}

$H_{in t}$ is given by
\begin{equation}
H_{i nt} = H_{xy}+H_{anis}
\end{equation}
\begin{equation}
\begin{array}{rl}
H_{xy} &= \aSU_{k,k^{\prime},q}
{\cal J}_{zz} (k,k^{\prime}, q)
\tau_{zkq} \tau_{zk^{\prime}q}  \nonumber \\
  &+ {\cal J}_{\perp} (k,k^{\prime},q)
( \tau_{xkq} \tau_{xk^{\prime}q} +
\tau_{ykq}
\tau_{yk^{\prime}q}) , 
\end{array}
\end{equation}
\begin{equation}
\begin{array}{rl}
H_{anis} &= \aSU_{k,k^{\prime},q} {\cal J}_{zx} (k,k^{\prime},q)
\tau_{xkq} \tau_{zk^{\prime}q} + h.c. \nonumber \\
&+ {\cal J}_{\perp}^{\prime} (k,k^{\prime},q)
( \tau_{xkq} \tau_{xk^{\prime}q} -
\tau_{ykq} \tau_{yk^{\prime}q} ) ,
\end{array}
\end{equation}
where
$\tau_{k,q}$ are defined in a-b space,
(not to be confused with momentum transforms of
(3.21)-(3.23)),
\begin{equation}
\begin{array}{rl}
\tau_{zkq} &= a_{k+q}^+ a_k -
b_{k+q}^+ b_k \nonumber \\
\tau_{xkq} &= \frac{1}{2} (a_{k+q}^+ b_k +
b_{k+q}^+ a_k ) \nonumber \\
\tau_{ykq} &= - \frac{i}{2} ( a_{k+q}^+ b_k -
b_{k+q}^+ a_k ) .
\end{array}
\end{equation}

As mentioned in Sec. 2, Eq. (4.1),
follows directly
from the ``bare Hamiltonian'', Eq. (2.1) by transforming to the
non-interacting bands using Eq. (2.5) 
(neglecting the terms in the
first line, which do not play a crucial role).
The transformation from the bare $U$'s and $V$'s to ${\cal J}$'s using (2.5) is
straightforward and not explicitly presented here.  
The only important point to note is that one should always include
both the particle hole channels.  Thus 
$V n_{di} n_{pj}$ is written as
\begin{equation}
\frac{V}{2} \left[ \sum_{\sigma , \sigma^{\prime}} 
(d_{i \sigma}^+ d_{i \sigma} ) 
(p_{j \sigma^{\prime}}^+ p_{j \sigma^{\prime}}) \;+\;
(d_{i \sigma}^+ p_{j \sigma^{\prime}})
(p_{j \sigma^{\prime}}^+ d_{i \sigma}) \right]
\;+\; \mbox{one-electron terms}
\end{equation}
We note that
the instabilities discussed here do not occur for the model with only
on-site interactions, just as in the case of the Hamiltonian derived
in the strong-coupling limit.
In the strong-coupling limit the kinetic energy parameters $\tilde{t}$
and $\Delta$ depend on the hole density $x$ through the
dependence of $\phi_0^2$ on $x$ derived below.
This in turn makes the effective interactions ${\cal J}$ depend on $x$ also.
If (4.1) is considered directly derived from (2.1), the kinetic energy and the interactions
are transformations of the bare terms.
One can interpret the operators $a_{k\sigma}$, $b_{k\sigma}$ as in
(3.32) and (3.33) or as the bare bandstructure operators given by
equation (2.5).

For short-range interactions, the ${\cal J}$'s can be
written as a sum over products of
separable functions with the symmetry
of the lattice.
In terms of the leading such terms, we define.
\begin{equation}
{\cal J}_{\eta \zeta} (k,k^{\prime} ,q) \equiv 
\sum_s {\cal J}_{\eta \zeta}
{\cal F}_{\eta}^s (k,q)
{\cal F}_{\zeta}^s
(k^{\prime},q) ,
\end{equation}
where $\eta , \zeta = (x,y,z)$.
The form-factors ${\cal F}_{\eta}$'s are obtained by
Fourier transforming (3.26) and using the rotations (3.32-3.33)
or directly from (2.1).

${\cal F}_{\eta^{\prime} s}$ have
a rather messy form.
I assume that there is one particular lattice harmonic which
dominates and henceforth drops the superscript $s$.  Of course,
the dominant harmonic can be determined only from a detailed
calculation.  Such calculations are not done in this paper, nor
are they necessary for the principal calculations drawn.

In (4.1) we have dropped the mean-field decomposition (3.31).
It simply renormalizes the kinetic energy in (4.1) without
introducing any new qualitative features.
The interactions term in (4.1) is purely in the spin-singlet
channel.
Due to lattice effects, coupling of the form
$\tau_{ykq} \tau_{zk^{\prime}q}$ and
$\tau_{ykq} \tau_{xk^{\prime}q}$ are also produced but they vanish as
$q \rightarrow 0$ and play no essential role.
I have dropped such terms.
\subsection{Instabilities}
We look for instabilities in the bandstructure of the $q=0$ intracell
excitonic nature due to $H_{i nt}$.
They can arise only if the parameters in $H_{i nt}$ are the same scale
as the bandwidth.
To calculate the properties of the new states, one introduces,
as usual uniform ($q =0)$ mean field amplitudes, which will
be determined variationally:
\begin{eqnarray}
T_z &\equiv& {\cal J}_{zz} \aSU_k
\langle \tau_{zk,0} \rangle 
{\cal F}_z (k) , \\
T_x &\equiv& \frac{1}{2}
{\cal J}_{xx}
\aSU_k
\langle \tau_{k,0}^+ +
\tau_{k,0}^- \rangle 
{\cal F}_x (k) ,  \\
T_y &\equiv& - \frac{i}{2} {\cal J}_{yy} \aSU_k
\langle \tau_{k,0}^+ -  \tau_{k,0}^- \rangle
{\cal F}_y (k)  \equiv \aSU_k
T_y (k) ,
\end{eqnarray}
where ${\cal F}_{\alpha} (k) \equiv {\cal F}_{\alpha} (k,0)$.
We also define the amplitude $T_{\perp}$
and angle $\theta$ by
\begin{equation}
T_{\perp} \equiv |T_x + iT_y | ,
tan \theta = T_y /T_x .
\end{equation}

The splitting of the bands always provides an effective
field acting on $\tau_z$.
Therefore any interesting instabilities can only be in the
$\tau_x - \tau_y$ plane.
We therefore look for instabilities which determine the
magnitude $T_{\perp}$ of an order parameter in the
$\tau_x - \tau_y$ plane and its angle $\theta$, with respect
to the $\tau_x$-axis.

If $H_{anis}$ were ignored, the model is isotropic about the z-axis.
There would be just one transition of second order nature with massless
collective fluctuations.
The coupling to Fermions of the collective modes would vanish
in the long wavelength limit because these modes arise
due to breaking a continuous symmetry.
$H_{anis}$ reduces the symmetry so that, as shown below, the general model
has one first-order transition and two second-order transitions of the
Ising variety as the parameters in the model, (including $\mu$) are
varied.

Before we proceed with the calculations, it is useful to discuss
the excitation spectra for interaction strength {\it less} than
necessary to cause the instability.
Consider first only $H_{xy}$.
The problem of the excitation spectra between a partially filled
band (a) and an empty (or fully filled) band (b) has been investigated
in degenerate semi-conductors$^{50}$ and with the approximation of a
dispersionless band (b) for the Fermi-edge singularities$^{50-52}$ in
the core spectra of metals.
The {\it absorption} spectra is given by, see fig. (9a).
\begin{equation}
\chi (\omega ,q) \sim
\aSU_{\nu}
\aSU_{k} \; \Lambda (k, \nu ; \omega , q)
G_a(k+ q, \nu+\omega )
G_b (k , \nu )
\end{equation}
where $\Lambda$ is the complete vertex in the particle-hole
channel with energy-momentum ($\omega , q)$,
and $G_a$ and $G_b$ are the (exact) single particle Green's
functions.
Using the fact that band $b$ is empty (at $T=0$), the sum over $\nu$
can be explicitly carried out with the result
\begin{equation}
\chi (\omega ,q) \sim
\aSU_{k} \Lambda (k,\nu ;\omega , q)
G_b (k, \nu ) \left |_{\nu=- \omega + \epsilon_{ak}- \mu} \right. .
\end{equation}
For small interactions, there is a modification of the spectra at the
{\it threshold} energy $\epsilon_t \equiv \epsilon_b (k_F)- \mu$.
We are interested only in interactions large enough
that an excitonic collective mode, which does not overlap the
inter-band $a$-$b$ transitions is pulled out.
The simplest approximation for the calculations is to consider a rigid
Fermi-sea which only serves to block out a part of the phase-space.
This is the ladder diagram approximation for $\Lambda$, fig. (9b).
In this case one obtains a sharp collective mode with spectral function
$\sim \; \delta (\omega - \omega_{ex} (q))$.
This is a poor approximation for the lineshape.$^{51}$
The dressing of the exciton by low-energy particle-hole excitations
at the Fermi-surface;
the simplest processes are represented in fig. (9c);
modifies the lineshape non-perturbatively.
The problem has been solved exactly$^{52}$
in the recoil-less limit, i.e. for a dispersionless b-band
where the interaction
strength can be parametrized by a phase-shift $\delta ( \epsilon )$.
In this case $\Lambda$ is also independent of momentum $k$ and
$\chi$ is therefore independent of $q$.
The absorption lineshape is as sketched in fig. (10a).
Near the excitonic threshold it is given by
\begin{equation}
\chi (\omega) \sim
( \omega-\omega_{ex} )^{-1+ \left ( 1- \frac{\delta_0}{\pi} \right )^2} ,
\omega > \omega_{ex} .
\end{equation}
Here $\delta_0$ is the phase-shift at the {\it chemical potential},
modulo $\pi$ -- the phase-shift required to pull an exciton
from the continuum.
$\omega_{ex}$ is determined by the details of bandstructure,
density of conduction electrons and the strength of the
potential.
Note that in the weak-coupling limit
the absorption lineshape has precisely the same
exponent,$^{51}$  but the absorption edge is at the energy $\epsilon_t$.
Thus Eq. (4.14) may be regarded as the Fermi-surface singularity pulled down
to $\omega_{ex}$, or that the absorption displays the excitonic edge
as modified by shake-off of low energy particle-hole excitations
at the Fermi-surface.
Therefore processes which smooth the Fermi-edge singularities
will also smooth the excitonic edge.

The effect of a finite hole mass or recoil$^{50}$ on the Fermi-edge spectra
is to smooth the singularity.
Auger-processes now introduce a self-energy
for $G_b$ which is smooth on the scale
of the recoil energy $\epsilon_b(k_F)- \epsilon_b$ (zone-boundary).
All $k$'s from the zone boundary to $k_F$ now contribute to the
absorption for any given external $q$, momentum is conserved by particle-hole
scattering on the Fermi-surface.
Both $G_b$ and $\Lambda$ are now functions of $k$.
The extra integrations in (4.12) then rounds off the singularity over
the recoil energy.
If interactions are strong enough to pull out an exciton, the
excitonic edge must be similarly rounded off around the
excitonic edge $\omega_{ex}$.
This is shown in fig. (10b).
Similar behavior must exist for a range of $q$
from 0 to order the difference from $k_F$ to the zone boundary.
We may write qualitatively that
\begin{equation}
\chi (\omega , q) =
\chi \left ( \frac{\omega -\omega_{ex} (q)}{\Gamma} \right ) ,
\end{equation}
where $\Gamma$ is the smaller of the recoil energy or $\omega_{ex}$.
$\chi$ $(x)$ has the form (4.14) for $x \gg 1$  but near
$x=0$, $ \chi (x)$ is a smooth function of $x$.

The two important points in the above discussion
are (i) that for large enough interactions, an excitonic
state is pulled out with or without recoil, (ii) that recoil
is always a relevant perturbation, smoothing out the singularity
at the Fermi-edge and therefore at the excitonic edge
if it exists.

As the interaction strength in (4.1) increase $\omega_{ex}(q)$ decreases.
The bandstructure
in Eq. (4.1) is unstable for interactions for which
$ Re \; \chi (0,0 ) \rightarrow \infty$.
It appears difficult to get explicit closed form
expressions for $\chi ( q , \omega )$ taking into
account the dressing of the exciton by low energy particle-hole pairs.
In subsequent sections, I present
explicit results with the frozen Fermi-sea approximation, and then
discuss from general considerations the essential features of the
exact $\chi ( \omega , q )$.

\subsection{Anisotropies}
The interband interaction terms ${\cal J}_{zx}$ and
${\cal J}_{\perp}^{\prime}$ play a quite different role than
${\cal J}_{\perp}$ and ${\cal J}_{zz}$.
It is convenient to first focus on these differences.
The effective interaction ${\cal J}_{zz}$
as derived in the appendix
is in general smaller than
${\cal J}_{xx}$ and ${\cal J}_{yy}$.
But the amplitude $T_z$ is always finite  because of the
splitting between the bonding-b and anti-bonding-a band in (4.1)
i.e. an effective uniform field $\sim \tau_z^i$.
The term proportional to ${\cal J}_{zx}$ in (4.4) is therefore approximated as
\begin{equation}
{\cal J}_{zx}^{-1} \left [ T_z T_x + T_z
\aSU_k \tau_{xk} {\cal F}_{xz} (k) +
T_x \aSU_k \tau_{zk} {\cal F}_{xz} (k) \right ]
\end{equation}
$T_z$ and $T_x$ will therefore have no low energy
dynamics.
So, at low energies, the corrections to the mean-field approximation (4.16)
are unimportant.
The second and third term in (4.16) merely renormalize the bandstructure
provided $T_z$ and $T_x$ are finite.$^{~~~}$
The first term 
may be written as
\begin{equation}
A cos \theta , \; \; A = {\cal J}_{zx}^{-1} T_zT_{\perp} .
\end{equation}

Consider next the term
proportional to ${\cal J}_{\perp}^{\prime}$
in (4.4).
In the mean-field approximation we write
it as
\begin{equation}
= ( {\cal J}_{\perp}^{\prime} )^{-1}
[ (T_x)^2 - (T_y)^2 ].
\end{equation}
Therefore this term acts as a quadratic anisotropy
field in the $x-y$ plane if an excitonic state condenses.
This anisotropy may be written as
\begin{equation}
B ( cos^2 \theta ) , \;  B = ({\cal J}_{\perp}^{\prime})^{-1} (T_{\perp})^2 .
\end{equation}

In general, higher anisotropies are also generated from the starting
Hamiltonian.
I do not give their derivation but merely introduce, in the
mean-field approximation a term, $C \; cos^4 \theta$ in the free-energy.
The mean-field anisotropy free-energy is then
\begin{equation}
F_{anis} = A \; cos \; \theta +
B \; cos^2 \theta + C \; cos^4 \theta .
\end{equation}
A, B and C are functions of the density of holes $x$ as well as, in general,
of temperature.
\subsection{Condition for Instability}

\noindent
{\bf (i) Rigid Fermi-sea}
As already mentioned only the ladder diagrams, fig. (8b) are considered
in this approximation.
\noindent

The mean-field free-energy then is
\begin{eqnarray}
F_{MF} =
\frac{T_z^2}{4{\cal J}_{zz}} &+&
\frac{T_{\perp}^{2}}{4{\cal J}_{\perp}}+
(E_{\phi} + \lambda ) \phi^2 - \lambda_0 \nonumber \\
    &-& \frac{1}{\beta} \aSU_{k,m=\alpha , \beta} ln (1-e^{-\beta (E_{km}-\mu )} ) 
    + F_{anis} .
\end{eqnarray}

We will be interested especially in the vicinity of the hole density
where the $F_{anis}$ vanishes to leading order.
It is convenient then to begin the analysis by ignoring
$F_{anis}$.
One thereby determines $T_{\perp}$, $T_z$ and $\phi$.
$F_{anis}$ is then used to determine $\theta$.

The variational one-electron Green's function is

\begin{equation}
G_0(k, \omega ) =
\left (
\begin{array}{cc}
G_{aa} & G_{ab} \\
G_{ba} & G_{bb}
\end{array}
\right )
\; = \;
( \omega - H_0 (k) )^{-1}
\end{equation}
\begin{equation}
H_0 (k) =
\left (
\begin{array}{cc}
\epsilon_{ka} + {\cal F}_z(k) T_z & {\cal F}_{\perp}(k) T_{\perp} \\
{\cal F}_{\perp} (k) T_{\perp} & \epsilon_{kb}
- {\cal F}_z(k) T_z
\end{array}
\right )
\end{equation}
The mean-field band structure $E_{k \; \alpha , \beta}$ and
the eigenvectors $\alpha_{k\sigma}$, $\beta_{k\sigma}$
are obtained as usual by diagonalizing
$H_0(k)$:
\begin{equation}
\left (
\begin{array}{cc}
\alpha_{k\sigma} \\
\beta_{k\sigma}
\end{array}
\right ) =
\left (
\begin{array}{cc}
c_k \; \; \; \; s_k \\
-s_k \; \; \; \;  c_k
\end{array}
\right )
\left (
\begin{array}{cc}
a_{k\sigma} \\
b_{k\sigma}
\end{array}
\right ) .
\end{equation}
Here
$$
\begin{array}{rl}
c_k &= cos \lambda_{k/2} ,
s_k= \; sin \; \lambda_{k/2}, \\
tan \; \lambda_k &= {\cal F}_{\perp} (k)
T_{\perp} / (\epsilon_{ka}- \epsilon_{kb}+ 2 \; {\cal F}_z (k)).
\end{array}
\eqno{(4.24a)}
$$
The leading term of ${\cal F}_{\perp} (k) = s_{xy} (k)$.

Minimizing $F_{MF}$ with respect to $\lambda$ and $\mu$ gives
\begin{equation}
\phi^2 + \aSU_{k,m} f (E_{km} -
\mu) = 1 ,
\end{equation}
and
\begin{equation}
\phi^2 = x ,
\end{equation}
where $f(z)$ is the Fermi-function.
Equation (4.25-4.26) imply that the Luttinger theorem on the
volume enclosed by the Fermi-surface is satisfied.

Minimizing $F_{MF}$ with respect to $T_z$, $T_{\perp}$ and $\phi$ yields
respectively
\begin{eqnarray}
\frac{T_z}{2{\cal J}_z} &+& \aSU_{km} f(E_{km} - \mu )
\frac{\partial E_{km}}{\partial T_z} = 0 \\
\frac{T_{\perp}}{2{\cal J}_{\perp}} &+& \aSU_{km} f(E_{km} - \mu )
\frac{\partial E_{km}}{\partial T_{\perp}} =0 \\
2\lambda \phi &+& \frac{\partial E}{d\phi} +
\aSU_{k,m} f (E_{km} - \mu )
\frac{\partial E_{km}}{\partial \phi} = 0 .
\end{eqnarray}
>From the appendix note that ${\cal J}_{\perp} \gg {\cal J}_z$.
We expect $T_z$ to have only a minor effect
which is determined mainly by the ``external field'' $\Delta$.
The stability of the mean-field approximation for $\phi_i$ is
ensured by a finite value for the ``Boson''
chemical potential $E_{\phi}+\lambda_0$.
In fact, apart from detailed quantitative issues, we need
look only at equation (4.28) which can be re-written as
\begin{equation}
\frac{1}{2{\cal J}_{\perp}} + \aSU_{k,\sigma}
|{\cal F}_{\perp} (k) |^2
\frac{f(E_{k\alpha \sigma}-\mu ) - f(E_{k\beta \sigma}-\mu )}{(E_{k\alpha \sigma}-E_{k\beta \sigma})} =0
\end{equation}
The (approximate) condition for one electron band structure to be unstable is
obtained by
\begin{equation}
\frac{1}{(2{\cal J}_{\perp})} + \aSU_{k\sigma}
| {\cal F}_{\perp} (k )^2
\frac{f(\epsilon_{ka\sigma}-\mu ) - f(\epsilon_{kb\sigma}-\mu )}{(\epsilon_{ka\sigma}-\epsilon_{kb\sigma})} =0
\end{equation}

At $T=0$, the right hand side of (4.31) is of order
$N(0) \; ln \; (W+ \epsilon_t)/ \epsilon_t$.
Since the ``threshold energy'' $\epsilon_t$ and the
bandwidth $W$ are of similar order, we need
$2 {\cal J}_{\perp} N(0)$ of O(1) to have an instability.
We will in fact assume that $2{\cal J}_{\perp} N(0)$ is large
enough that the instability is at a very high temperature.
Stability is achieved by $T_{\perp} \neq 0$, which
corresponds simply to changing the relative Cu and O character of
the occupied and unoccupied band.
This result is no more than the statement that just as
at $1/2$ filling the charge state away form  $1/2$ filling is
determined by the electron-electron interactions, not just
by the one-electron bandstructure.
This behavior has been seen in a variety of earlier calculations.$^{41,42}$
We will see below that when $F_{anis}$ is considered, the transition
to $T_{\perp} \neq 0$ becomes of first order,
except at two points.
It is only near those two points that interesting properties can arise.

\noindent
{\bf (ii) Soft Fermi-sea}

\noindent
Only the ladder diagrams are considered in the vertex $\Lambda$ in deriving
Eqs. (4.27)-(4.29); i.e. the Fermi-sea merely acts to restrict phase-space.
As already discussed, this is a poor approximation.
For sufficiently strong interactions the one-electron band structure is
of course unstable even with the inclusion of low-energy particle-hole
excitations, such as in Fig. (8c,d).
But several details change.
In general, the mean-field amplitudes $T_x$, $T_y$ and $T_z$ are functions
of frequency.
But as a variational ansatz, Eqs. (4.8)-(4.10)
may still be introduced.
The considerations of anisotropy still continue to hold as in Sec (4.b).
The mean-field free-energy (4.21) is an approximation of the general case,
where the $T$ and $\phi$ dependence (after integrating over the
Fermions) may be written as
\begin{equation}
\left ( T_{\perp} \; T_z \; \phi \right )  \; \chi^{-1}
\left (
\begin{array}{cc}
T_{\perp} \\
T_z \\
\phi
\end{array}
\right ) .
\end{equation}
Equations (4.27)-(4.29) are rigid Fermi-sea approximation of
the general condition
\begin{equation}
det \; \chi^{-1} (\omega=0, \; q=0) =0
\end{equation}
to determine the variational parameters
$T_{\perp}$, $T_z$ and $\phi$.

The qualitative form of $\chi ( \omega ,q)$ for
interaction strength less than the critical
value has been discussed in Sec. (4a) and
illustrated in fig. (10b).
Let us define $p$ as the parameter (which is
a function of the parameters in the Hamiltonian,
(4.1)) such that the instability towards
$T_{\perp} \neq 0$ occurs at a temperature
$T_c$ for $p=p_c(T_c)$.
First consider $T_c \approx 0$.
What does the condition (4.33) for the instability imply for
$Im \; \chi ( \omega ,q)$ when the
latter is overdamped and has the shape  as in fig. (10b) rather than a delta-function
as in the rigid Fermi-sea approximation.
$\chi (\omega ,q)$ has to satisfy the requirement
\begin{equation}
Im \; \chi (\omega , q) = - Im \; \chi (- \omega ,q).
\end{equation}
For $\omega$ small compared to $\omega_{ex}$, $Im \; \chi (\omega ,0) \sim \omega$ while
for $\omega$ large compared to $\omega_{ex}$, it is nearly a
constant up to a cutoff $\omega_c$ on the scale of the
Fermi-energy.
Then by Kramers-Kronig transform the leading term in
\[
Re \; \chi (\omega , 0 ) \; \sim \;
ln \; \left ( \frac{\omega_c}{max (\omega_{ex}, \omega)} \right )
\]
As $p \rightarrow p_c(0)$, $\omega_{ex} \rightarrow 0$.
Near this point, $Im \; \chi (\omega ,0) \sim sgn (\omega )$.

We see that recoil
reduces the singularity as $\omega \rightarrow 0$ of
$\chi (\omega , 0)$ near the transition at $T=0$ as $p \rightarrow p_c(0)$
from the $\delta$-function of the rigid sea approximation or the
exact result (4.14) for the recoil less case.
In terms of (4.14) recoil makes the phase-shift $\delta_0$ at the
chemical potential irrelevant.
This appears quite unavoidable -- on the one hand recoil cannot
prevent the instability if interaction is large enough, on
the other hand the singularity in (4.14) which is a Fermi-edge
singularity is wiped out by recoil.
The result is that the instability occurs with the
least singular form possible:
$\chi (\omega , 0) \sim ln (\omega + i0)$.
As discussed in connection with Eq. (4.15),
the smoothing of the excitonic edge occurs at any external $q$ due
to the mixing by Auger processes of inter-band particle-hole
pairs over essentially the whole range of momenta.
Thus for interaction energies large compared to the Fermi-energy
as required for the instability
and therefore also large compared to the recoil energy,
the frequency dependence of the
$Im \; \chi ( \omega ,q)$ is nearly the same over
the whole range of $q$.
The imaginary part of $\chi (\omega ,q)$ is then reminiscent
of the form of the Cooper pair-fluctuation propagator$^{53}$ above
$T_c$ which is $\sim i \omega / max (\omega , T)$ over a range of $q$
smaller than the coherence length $\xi_0$, i.e. the size of the
Cooper pairs.
Here, given the strong coupling required to engender the instability,
the excitons have size of the order of the lattice spacing.
So the $i\omega / max ( \omega ,T)$ form of damping is expected to
persist over most of the Brillouin zone.

The form for $\chi (\omega , q)$ near
for $T=0$ as $p \rightarrow p_c(0)$ is thus
\begin{equation}
\chi (\omega , q) \sim
\left [ \left ( \frac{i \; \omega}{max ( \omega , \omega_{ex} (p))} +
ln \; \frac{\omega_c}{max ( \omega , \omega_{ex} (p))} \right )^{-1} +
\kappa^2 q^2 +
(p_c(0) - p) \right ]^{-1} .
\end{equation}
where $\omega_c$ is an upper cut off energy,
and $\kappa$ provides the scale of dispersion.
At a finite temperature, we must use
the fact that
the $\omega$ and $T$
dependence in $\chi( \omega , q )$ must scale as $\omega / T$.
So, for finite $T \gg \omega$
\begin{equation}
\chi ( \omega , q , T) \sim
\left [ \left ( \frac{i \omega}{max ( T, \omega_{ex} (p))} +
ln \; \frac{\omega_c}{max (T, \omega_{ex} (p))} \right)^{-1}
+ \kappa^2 q^2 + (p_c (T) -p) \right ]^{-1}
\end{equation}

\subsection{Determining $\theta$}
We now consider the effects of $F_{anis}$.
On minimizing $F_{anis}$ with respect to $\theta$, one
finds that the equilibrium value $\Theta_m$ is given by
\begin{equation}
\mbox{Phase} \; \; {\bf I}: \; \;
\Theta_m = 0 \; \; \; \mbox{for} \; \; \; (A+2B+4C) < 0
\end{equation}
and
\begin{equation}
\mbox{Phase} \; \; {\bf I}^{\prime}: \; \;
\Theta_m =\pi \; \; \; \mbox{for} \; \; \; (-A+2B+4C) < 0 .
\end{equation}

For $C > 0$, there occurs a second order transition of the Ising variety to
\begin{equation}
\mbox{Phase} \; \; {\bf II}: \; \; \;
0 < \Theta_m < \pi , \; \; \mbox{for} \; \; \;
-A \langle 2B+4C < A .
\end{equation}
$\Theta$ continuously rotates in phase II as A, B, C vary.

Noting that A, B, C are in general functions of $x$ and $T$, we may write the mean-field anisotropy energy as
\begin{equation}
F_{anis} = G_0 (x,T)
(\theta - \Theta_m (x,t))^2 + ... .
\end{equation}
where
\begin{equation}
G_0(x,T) =0 , \; \; \mbox{for} \; \; \pm A+2B+
4C=0, \;\; \mbox{i.e.} \; \;  \mbox{at} \; \;\Theta_m = 0, \pi .
\end{equation}
Therefore the transition at $\Theta_m=0$ or $\pi$
is of second order.
Let us denote the transition line (understanding that
as $x$ is varied
we will be concerned only with either
$I$ to $II$ or $I^{\prime}$ to $II$ transition) by $x_c(T_c)$.

Let us stay for definiteness in the vicinity of
$\Theta_m=0$.
For $\Theta_m=0$, the high temperature transition occurs as a first order transition with
a real order parameter $T_x$.
When $\Theta_m \neq 0$, the mean-field order parameter is complex;
$T_x+i \; T_y$.
$T_y \neq 0$ implies that in the ground state, a current
distribution occurs within each unit cell which has the same phase in every unit cell.

The content of the mean-field theory is summarized in fig. (5),
where the free-energy is shown as a function of $T_x$ and $T_y$.
Throughout the temperature region of interest
$T_x \neq 0$.
As ($p \equiv \pm A + 2B+4C$) is varied by varying $x$ (and $T$), a second
order Ising transition from $T_y=0$ to
the circulating current phase $T_y=\neq 0$ occurs.
(We note in passing that we have arrived at a new class of
statistical mechanical model for quantum-critical points.)

We will discuss below how regimes 1 and 3 of the phase
diagram of fig. (1) may be identified with the Phase I (or $I^{\prime}$)
and regime 2 and 4  with the Phase II of the mean-field theory.
The nature of the fluctuations will be shown to vary in Phase I
as a function of $x$ and $T$ leading to a cross-over in the
properties from regime 1 to regime 3 in fig. (1).
We will also show that the phase transitions between Phase I and
Phase II also becomes a cross-over for arbitrarily small disorder.
\subsection{The Circulating Current Phase:}
$\Theta_m \neq ( 0, \pi )$ implies that in the ground state, a
current flows in each cell.  Since the momentum of the instability
is zero, the current pattern respects lattice-translation
symmetry.  The current pattern within a cell can be deduced from
the mean-field Hamiltonian with $\Theta \neq 0$, which is now
Eqn. (4.23) with the substitution 
$T_{\perp} \rightarrow T_{\perp} e^{i \Theta}, T_{\perp} e^{-i
\Theta}$ in the off-diagonal terms.  To calculate the current
pattern, first find the eigen-vectors of the conduction band with
$\Theta \neq 0$:
\begin{equation}
E_{\alpha} (k) \hat{\alpha}_{k \sigma} =
( \epsilon_{ka} + {\cal F}_z (k) T_z )
a_{k \sigma} + {\cal F}_{\perp} (k) T_{\perp} e^{i \Theta}
b_{k \sigma}
\end{equation}
Now use Eqs. (2.5) to express $a,b$ in terms of $d_k$ and
$p_{x,y} (k)$ using coefficients given in Appendix A.  This yields
\begin{equation}
\hat{\alpha}_{k \sigma} = \hat{u}_{\alpha d} ({\bf k}) 
d_{{\bf k} \sigma} + \hat{u}_{\alpha x}
({\bf k}) p_{x {\bf k} \sigma} 
+ \hat{u}_{\alpha y} ({\bf k}) p_{yk \sigma}
\end{equation}
where $\hat{u}$'s are complex coefficients
\begin{equation}
\hat{u}_{\alpha (d,x, y)} (k) \equiv R_{\alpha (d,x,y)}
(k) e^{i \phi_{(d,x,y)} (k)} .
\end{equation}
There is no need to exhibit the complicated expressions for these
coefficients because the current pattern can be deduced from their
general properties specified below.

The current in a bond going from a copper site to an oxygen site
in the $x$ or $y$ direction is 
\begin{equation}
\begin{array}{rl}
j_{dx}
= & \frac{2}{\pi} \; t_{pd} \;Im \; \langle d_{i \sigma}^+ p_{i+x, \sigma}
\rangle \\
= & \frac{2a}{\pi} \; t_{pd} \sum_{k < k_F} cos \;
\frac{k_x a}{2} \; R_{\hat{\alpha} d} \; R_{\hat{\alpha} x}
sin ( \phi_d (k) - \phi_x (k)
\end{array}
\end{equation}
(The sum over $k$ of the term with $sin \frac{k_xa}{2}$
is zero using inversion symmetry.)  Now note that $t_{pd}$
changes sign $x \rightarrow - x$ and the sum in (4.45) is
symmetric under inversion.  Hence the current between a
copper-orbital at a site $i$ and an oxygen orbitals at
$i + \frac{a}{2} \hat{x}$ and $i - \frac{a}{2} \hat{x}$ are equal
and opposite.  This holds also for the current between
copper-orbitals at $i$ and oxygen orbitals at 
$i \pm \frac{a}{2} \hat{y}$.

The current between two oxygen orbitals in the same cell
\begin{equation}
\begin{array}{rl}
j_{xy}
& = \frac{at_{pp}}{a} \; Im \langle p_{i+x, \sigma}^+
p_{i+y, \sigma} \rangle \\
& = \frac{2at_{pp}}{a} \; \sum_{{\bf k}<k_F} cos\;
\frac{k_{xa}-k_{ya}}{2} \;
R_{\hat{\alpha} x} ({\bf k}) \; R_{\hat{\alpha} y} ({\bf k}) \; sin
(\phi_x ({\bf k}) - \phi_y ({\bf k}) )
\end{array}
\end{equation}
The sum in (4.46) is identical for the other three oxygen-oxygen
bonds around a given copper atom, but as is evident from the
phases shown in Fig. (8) $t_{pp}$ reverses sign cyclically in
going around the four bonds, and therefore so does the current.

The direction of the current between the copper and the oxygen
orbitals and between the oxygen orbitals fixes the pattern show in
Fig. (6).  So together with breaking time-reversal symmetry,
four-fold rotational symmetry is broken.  But the product of the
two is left invariant.

Some further conclusions can be drawn from an examination of the 
$\hat{u}$'s.  If $t_{pp} = 0$, $\psi_{\alpha x} ({\bf k}) =
\psi_{\alpha y} ({\bf k})$.  Then any
$( \psi_{\alpha d} (k) - \psi_{\alpha x,y} (k))$ can be removed by
a unitary transformation without affecting the eigenvalues or the
eigen-vectors.  This is physically obvious from looking at Fig.
(6); a current between copper and oxygen orbitals is meaningless
in the absence of a current between the oxygen orbitals.

We can also deduce that there is no contribution to the currents
from states on the diagonals in the Brillouin-zone,
$\pm k_x = \pm k_y$.  Correspondingly, there is no change in the
single-particle eigenvalues in the circulating current phase along
the diagonals.  On zone faces
$k_x = 0$ or $\pi /a$, $k_y = 0$ or $\pi / a$, the eigenvalues do
change.  The lowest lattice harmonic consistent with these
symmetries is $d_{x^2 - y^2}$.  So there is a change in the single
particle spectra in the circulating current phase of
$d_{x^2 - y^2}$ symmetry.

The ground state current contribution of each state ${\bf k}$
depends on ${\bf k}$ and there is $O (1)$ electron per unit cell
in the conduction band.  The orbital magnetic moment of the
circulating current in each of the quadrants in Fig. (6) is
$O (0.05 \mu_B )$, with the assumption that the average state
contributes $\sim 1/4 \mu_B$ per unit cell.

\section{Collective Modes and Fermion-Boson Coupling}
\setcounter{equation}{0}
We now consider the fluctuations in the state
$T_{\perp} \neq 0$.
They are interesting only near the $I$ (or $I^{\prime}$) to $II$ transition.
There is always a finite effective field coupling linearly to $T_z$.
So the
fluctuations in the z-direction are
always massive.
The interesting modes are in the $T_x-T_y$ space.
So define
\setcounter{equation}{0}
\begin{eqnarray}
\delta T_{x,q} &=& \frac{1}{2} {\cal J}_{xx}
\aSU_k \langle \tau_{k,q}^+ +
\tau_{k,q}^- \rangle {\cal F}_x (k,q) -
T_x \\
\delta T_{y,q} &=& - \frac{i}{2}
{\cal J}_{yy} \aSU_k \langle \tau_{k,q}^+ -
\tau_{k,q}^- \rangle {\cal F}_y (k,q) - T_y .
\end{eqnarray}
The effective Hamiltonian determining the fluctuations is
\[
H_{fluc} = \aSU_{k\sigma}
\left ( a_{k\sigma}^+ \; \; b_{k\sigma}^+ \right ) H_0 (k)
\left (
\begin{array}{c}
a_{k\sigma} \\
b_{k\sigma}
\end{array}
\right ) +
\aSU_q \frac{1}{4{\cal J}_{\perp}} ( \delta T_{x,q}^+
\delta T_{x,q} + \delta T_{y,q}^+ \delta T_{y,q} ) 
\]
\[
\begin{array}{ll}
&+ \aSU_{k,q} {\cal F}_{\perp} (kq)
(a_{k+q}^+ b_k + b_{k+q}^+ a_k )
( \delta T_{x,q} + \delta T_{x-q}^+ ) \\
&+ i \aSU_{k,q} {\cal F}_{\perp} (kq)
(a_{k+q}^+ b_k - b_{k+q}^+ a_k)
(\delta T_{y,q}+ \delta T_{y,-q}^+ ) .
\end{array}
\]
\begin{equation}
+ H_{anis} .
\end{equation}

Again, let us ignore the effects of anisotropy to  begin with
but choose $\Theta_m=0$, i.e. $T_{\perp} = T_{x}$.
The spectrum of the fluctuations
\begin{eqnarray}
D_{x}^0 (q,\omega ) &\equiv& \langle \delta T_x \delta T_x \rangle (q,\omega ) \\
D_y^0 (q,\omega )  & \equiv&
\langle \delta T_y \delta T_y \rangle (q,\omega )
\end{eqnarray}
is given in the frozen Fermi-sea approximation,
(fig. 8b), by
\begin{equation}
\left\{
\begin{array}{rl}
D_{x}^{0-1} (q, \omega ) \\
D_{y}^{0-1} (q, \omega )
\end{array}
\right\}
\begin{array}{rl}
= \frac{1}{2{\cal J}_{\perp}} &+ \aSU_{k,\nu}
| {\cal F} (k,q) |^2 
[G_{aa} (k+q,\omega + \nu) G_{bb} (k,\nu) \\ \nonumber
         &\pm G_{ab} (k+q,\omega +\nu) G_{ba}(k,\nu) ].
\end{array}
\end{equation}
At $\omega=0$, $q \rightarrow 0$, the equation for $D_{y}^0$
is identical to Eqn. (4.30) determining $T_x$.
So a long wavelength massless phase or current mode exists as is
to be expected when the anisotropy in the
${\tau}_x - {\tau}_y$ plane is zero.
The poles of $D_x^0$ give the dispersion of the
amplitude modes.
Their frequencies near $q \approx 0$ are of order $T_x$;
they will not be considered further.

Let us now include the effect of anisotropy, but stay near
{\bf I} to {\bf II} (or
${\bf I}^{\prime}$ to {\bf II}) instability.
>From (4.40), the anisotropy energy provides a
quadratic term $G_0(x,T) \delta T_y^2$ to the
fluctuations.
Including this effect
\begin{equation}
D_{y}^0 ( {\bf q} ,\omega )=
\frac{m/m^*}{\omega^2 - \kappa^2 q^2+G_0 (x,T)}
\end{equation}
where $\kappa /a$ is the order of $T_x$.
The spectral weight of the collective mode
$m^*/m$ is
\begin{equation}
\frac{m^*}{m} \sim
0 (T_x / W).
\end{equation}

Equations (5.6) are special cases, in the rigid
Fermi-sea approximation of the general equation
\begin{equation}
\chi^{-1} (q, \omega ) = 0
\end{equation}
which determines the fluctuation spectra, just as 4.27-4.29 are
special cases of Eqn. (4.33) which determines the
one-particle spectra through fixing $T_x$ etc.
In the frozen Fermi-sea approximation, there is no damping
of the collective fluctuations -- the excitonic resonances
have a spectral function proportional to a $\delta$-function.
As discussed in Sec. (4a) for the case when $T_x =0$, inclusion of low-energy
particle-hole fluctuations change the spectral function of the
excitonic collective mode in an essential way.
The $\chi(q, \omega )$ including low-energy particle-hole fluctuations
with $T_x \neq 0$
has the same functional form as discussed in (4c), but
calculated with the new bandstructure,
Eqns. (4.23), (4.24).
The dispersion of the soft excitonic
collective mode at $q \rightarrow 0$, described by $D_y^0 (q, \omega )$,
again has the same form for $G=0$
as discussed in Sec. (4c),
leading to Equations (4.35) and (4.36).
Including the effect of the anisotropy on the fluctuations, we have
\begin{equation}
D_y^0 (q, \omega ) = D_0
\left [ \left \{ \frac{i \omega }{max ( | \omega | , T,G_0)}  +
ln \left (  \frac{ \omega_c}{max ( | \omega | , T,G_0 )} \right ) \right \}^{-1} +
\kappa^2 q^2 + G_0 (x,T) \right ]^{-1} .
\end{equation}
Here $D_0$ parameterizes the spectral weight of the fluctuation
expected to be of $O(E_F^{-1})$.
The second order transition occurs when $G_0(x,T) =0$ as in
Eqs. (4.37) - (4.39).
(Henceforth $G_0$ is dimensionless,
having been scaled by $D_0^{-1}$).
The fluctuations have a finite frequency $G_0$ above and
a finite value below the transition
(characteristic of transitions of the Ising class).

Equation (5.10) is crucial in the analysis below of the properties
of the model.
It is clear that the very singular result from the rigid
Fermi-sea approximation, Eqn. (5.7) is quite incorrect.
(It also gives properties in $d=2$ which are too singular
compared to experiment.)
The combined effect of infrared processes at the Fermi-surface
and recoil together with analyticity requirements has been discussed in Sec. (4c) to lead to (5.10).
This justification is only heuristic.
An evaluation of processes like in fig. (9c) to find $\Lambda$ exactly  appears
very hard, if not impossible.
Earlier, a three-body scattering approach to the problem was
suggested.$^{54}$
It might be possible to evaluate $D_y(q, \omega )$ systematically
in such an approximation.
Note that when $\delta_0$ of Eq. (4.14) is zero, as
argued here, $Im \; G_b (\omega ) \sim \omega^{-1}$,
at least for the recoilless case.$^{50}$
This is consistent with the conjecture$^{54}$ that a three-body
resonance at the chemical potential may lead to
the observed normal state anomalies.

There are no other massless modes in the model.
Earlier investigations$^{41,42}$ of the charge
transfer instability in the model found a diverging
compressibility indicating phase-separation.
If one adopts a short-range interaction model, the density fluctuations
have a dispersion $\omega \sim q$.
Near the critical point of the charge transfer instability,
a low energy mode of $\delta T_x$ couples to such density
fluctuations pushing the ``electron sound velocity''
to zero.
In a model with Coulomb interactions, the density fluctuations are
at the plasma frequency.
Phase separation then does not occur (unless the system has inhomogeneously
distributed fixed (ionic) charges).

Consider now the coupling of the Fermions to
the low-energy collective modes in the vicinity of $x_c(0)$ where a transition
from $T_{\perp}=T_x$ to a {\it complex}
order parameter $T_x+i T_y$ occurs.
The coupling of the Fermions to the $\delta T_y$ fluctuations
comes from the fourth term in (5.3) and a similar term in (4.4).
We must re-express the Fermion operators in terms of the low-energy
Fermions created by $\alpha_{k \sigma}^+$ by using the rotation (4.24).
The coupling is written as
\begin{equation}
H_{F-B} = \aSU_{k,q,\sigma}
i \; g(k,q) \; \alpha_{k+q,\sigma}^+
\alpha_{k,\sigma} (\delta T_{yq} +
\delta T_{y-q}^+ ) .
\end{equation}
One finds from the fourth term in (5.3), (corresponding term
from (4.4) introduces no important difference)
\begin{equation}
g(k,q) = g_0  \; {\cal F}_y
(k,q) \; sin \; \left ( \frac{\lambda_{k+q} - \lambda_k}{2} \right ) .
\end{equation}

In (5.12) we have introduced $g_0$ with dimensions of energy,
so that $\delta T_y$ is henceforth dimensionless.
$g_0$ is expected to be of $O(E_F)$.
In (5.12) ${\cal F}_{y}$ is given by (4.7).
$g(k,q)$ vanishes linearly with $q$.
So even through we have an Ising transition, the coupling of the Fermion to the
fluctuations vanishes at long wavelength.
This occurs because the bands are split due to
``external'' fields and the fact that no $\tau_x \tau_y$ or
$\tau_z \tau_y$ coupling is allowed at long wavelengths.
The vanishing of the coupling at long wavelengths is however not
fatal because in the fluctuation spectrum Eqn. (5.10), all
$q$'s are (to a logarithmic accuracy) equally important in properties
which integrate the spectrum to energies of the order of temperature.
\section{Analysis of the Low-Energy Hamiltonian}
\setcounter{equation}{0}
The end result of the preceding two sections is a
simple Hamiltonian for the calculation of low energy properties:
\begin{equation}
\begin{array}{rl}
H &= \aSU_{k,\sigma} \epsilon (k)
\alpha_{k\sigma}^+ \alpha_{k\sigma} + D_y^{0-1} (q, \omega)
\delta T_{y,q}^+ \delta T_{y,q} \\ \nonumber
  &+ \aSU_{k,q} i g (k,q)
\alpha_{k+q\sigma}^+ \alpha_{k,\sigma}
(\delta T_{y,q} + \delta T_{y,-q}^+ ) .
\end{array}
\end{equation}
We now analyze the properties of this Hamiltonian in the regime near
the I to II transition near $T=0$.

It is best to start by considering the simple physical
processes depicted in figs. (11a-f).
Here the wiggly lines denote the fluctuation propagation
$D_y^0(q,\omega )$ and the solid lines the Fermion propagation
$G_0 (k,\omega) = \langle \alpha \alpha^+ \rangle (k,\omega )$.

The real part of 11(a) for $q \rightarrow 0$, $\omega \rightarrow 0$
renormalizes the mass $G$.
This can be absorbed in a redefinition of $x_c$.
Similarly the leading $q$ dependence $\sim q^2$ merely redefines
the coefficient $\kappa$ in Eqn. (5.10).
The imaginary part gives the usual Landau damping contribution
$\sim \; i \omega / v_F q$.
If we use renormalized Fermion propagators in fig. (11a),
the result is modified to
\[
i \omega / max ( Im \; \Sigma (\omega , T,G),
v_Fq).
\]
We will show that in the non-Fermi-liquid regime
$Im \; \Sigma (\omega , T) \sim max (\omega ,T)$.
This is an additive correction to the imaginary part in (5.10) and
is therefore unimportant at all $q$.
This would not be true in the rigid Fermi-sea approximation in which
the fluctuations in $D^0 (q, \omega )$ are completely undamped.
(In that case $D^0$ has the same functional dependence on $q$ and
$\omega$ as the transverse electromagnetic field
propagators in a metal.)$^{55}$
One can also examine higher order renormalizations, Fig. (11b) and
fig. (11c) to conclude that they are irrelevant, the imaginary part
of the process in (11b) is proportional to $\omega$ while (10c)
is proportional to $\omega^2$.

Consider next the Fermion self-energy graph, fig. (11d).
The imaginary part of the self-energy is easily seen to be
\begin{eqnarray}
Im \; \Sigma (q,\omega ) &=&  N(0) \int_0^1
d^{d-1} x
\int_{\epsilon_1}^{\epsilon_2}
d \epsilon \; Im D_R^0
(2 k_F x, \omega - \epsilon ) g^2 (2k_Fx)  \\ \nonumber
 &\times& \left [ tanh \; \frac{\epsilon}{2T} +
coth \; \frac{\omega - \epsilon}{2T} \right ] .
\end{eqnarray}
where
\begin{equation}
\epsilon_{1,2} = v_F (|q| \mp k_F x )+ Re \; \Sigma (\epsilon_{1,2} ) .
\end{equation}
It is found consistent to ignore $\Sigma (\epsilon )$ in the
right-hand side of (6.3).
For $T=0$, the limits of $\epsilon$ integration are 0 to $\omega$
(except for an ignorable region of $x$ integration of
$0(\omega / E_F)$).
The $x$-integration, using (5.10) for $D_0$ for $G_0 =0$ then leads to a
constant, so that the final result is proportional to $\omega \; sgn \; \omega$.
Similarly for $T \ll \omega$, the result is proportional to $T \; sgn \; \omega$.
The self-energy has negligible momentum dependence (if the Fermi-surface
has no significant nesting).
These results are true for any dimension more than 1.
The $d$-independence of the self-energy, and other properties in
which these fluctuations are sampled over energies to the
scale of the external frequency and temperature arises because in (5.10),
the fluctuations are essentially local (even though $Re \; D(0,0)$
diverges as $G^{-1}$).
We may express both the real part and the imaginary part of the
self-energy by an expression which interpolates between the
$\omega / T \ll 1$ and the $\omega / T \gg 1$ limits:
\begin{equation}
\Sigma (\omega , T) = \pi \lambda
\left [ (2\omega / \pi ) ln
\left ( \frac{\pi T+i\omega}{\omega_c} \right ) +
i \pi T \right ] .
\end{equation}
Equation (6.4) may be useful in analyzing angle-resolved photoemission experiments
discussed below.

Consider next the vertex correction shown in graph (10e).
In the limit
$q \rightarrow 0$ first and then $\omega \rightarrow 0$,
it is given by a Ward-identity
(in the pure limit),
\begin{equation}
\Lambda^{\omega} = \frac{1}{z}
\end{equation}
where $z$ is the quasiparticle renormalization amplitude
given by (1.2).
In the ``$q$-limit'' it is given by another
Ward-identity in terms of $d \Sigma / \partial k$.
Since $\Sigma$ is found very weakly dependent on
($k-k_F$), this is ignorable.

For general $\omega$ and $q$, a finite vertex correction O(g)
non-singular as a function of $\omega$ and $k-k_F$)
is found.
If the ``bare'' coupling constant $g(k,q)$ is less than O(1),
this may be simply absorbed in the redefinition of $g(k,q)$.
One can formally devise $\frac{1}{N}$ schemes to keep such vertex corrections
controlled.

We briefly consider the renormalization of $G_0$ in
$D_0 (q, \omega )$ due to anharmonic interactions.
The leading contribution comes from
\begin{equation}
u \; | \delta T_{q,\omega} |^2 \;
| \delta T_{q^{\prime}} , \omega^{\prime} |^2
\end{equation}
where $u > 0$ is a phenomenological coefficient expected to be
on the same scale as the upper cut off energy of the fluctuations.
The self-energy of the modes proportional to $u$, fig. (11f)
has the leading temperature correction proportion to $uT$
(independent of $d$) for $x \approx x_c(0)$.
This may be absorbed in $x_c(T)$) and suppresses the transition temperature.
(Hence forth we can drop the superscript 0 on $G$ and $D$.)
A proper analysis of the fluctuations near the transition line
which changes from a quantum transition at $T=0$ to a classical
Ising transition at high temperature has not been carried out.

A very important general point to note in this connection is that
the correlation length exponent $\nu$ at $T=0$ as a function of
$(x-x_c)$ is 0 while it is 1 for the $d=2$ classical Ising model.
This is expected to turn the transition line to a cross-over
for arbitrarily small disorder as discussed in Sec. (8).

We can estimate roughly the different regimes of fluctuations from the
properties of the propagator $D(q,\omega )$.
At this point no sophisticated analysis of the crossover between
different regimes is attempted.
Consider first the departure of the transition temperature of the
circulating current phase from $T=0$ at $x=x_c(0)$.
We assume parameters are such that at $T=0$, the CC phase occurs for
$x \leq x_c(0)$.
This is provided by an estimate based on the parameters
calculated in the appendix and the decrease of $\Delta$ with $x$.
Then the transition temperature $T_c(x)$, given by the divergence in
$Re \; D(0,0)$ is at
\begin{equation}
\left ( ln \; \left | \frac{\omega_c}{max(T_c, G )} \right | \right )^{-1}
= -G  \; \mbox{for} \; G < 0.
\end{equation}
This gives
\begin{equation}
T_c \simeq - G(T_c)
\end{equation}
i.e., the transition temperature is essentially
proportional to $- G(0)$.

For finite $T_c(G )$, in a narrow temperature $\Delta T_c$ region
near $T_c (G )$, the fluctuations are characteristic of the
classical Ising model.
We have not investigated here how the width of this regime
varies with $T_c$.
In this regime, the classical thermal occupation of fluctuations (where
$\frac{1}{2} coth ( \omega / 2T ) = ( n ( \omega / T) + \frac{1}{2} )  \approx \frac{T}{\omega}$),
determines the thermodynamic and
other properties because the characteristic energy of the fluctuations
is $O(\Delta T_c ) \ll T$.
If the characteristic energy of the fluctuations is much larger
than $T$, the zero-point occupation of the fluctuations
dominates the properties, (in this regime $coth \omega / T \approx 1$).
The physical properties in this regime are
governed by the quantum fluctuations.
Within this regime, we must distinguish when the characteristic scale
of the fluctuations is given by temperature itself and when it is
given by $G(x)$.
The former is the non-Fermi-liquid regime and the latter the Fermi-liquid
regime.
>From the form of $D(q, \omega )$, the cross-over between the two
occurs at $T \approx G (x)$ for $G > 0$.

The momentum integrated fluctuation spectrum for
$T \gg | G |$ gives a measure of the
frequency distribution:
\begin{equation}
\begin{array}{ll}
\int d^2 q \: Im D(q,\omega ) &\approx \frac{1}{\pi}
\left ( \pi / 2 - tan^{-1}
\left [ \frac{max(\omega , T)}{\omega}
\left ( ln \: \frac{\omega_c}{max (\omega , T)} \right )^{-1} \right ] \right ) \\
 &\sim \frac{\omega}{T} \: ln \: \frac{\omega_c}{T} \; \; \; \mbox{for} \;\;
\omega \ll T \\
 &\sim ln \: \frac{\omega_c}{\omega} , \; \; \; \; \mbox{for} \;\; \omega \gg T .
\end{array}
\end{equation}
Above the narrow critical regime near $T_c( G )$, the
properties are governed by the quantum fluctuations.
Non-Fermi-liquid behavior is to be expected.
For $G < 0$, i.e. the ordered side, the fluctuations have a gap.
So Fermi-liquid behavior (but with unusually small parameters) is to
be expected in the pure limit.
We will discuss below that this regime is very sensitive to disorder.

For $G > 0$, the fluctuations have a gap of $O(G )$ for
$T \ll G$.
Fermi-liquid behavior is therefore to be expected but with parameters
determined by $G$.

The different regimes are shown in fig. (12).
We are now ready to calculate the physical properties of (6.1).
Given (6.1), all physical properties can be calculated in a controlled
and systematic fashion because of the unimportance of
vertex corrections in the Fermion-Boson scattering and the Boson propagator.
For instance the single particle self-energy may be calculated
self-consistently by using the renormalized Fermion propagator in (11a).
The answer remains unaltered as in other problems with
momentum independent self-energy.
In the fluctuation spectra (5.10), all momenta are
equally important to logarithmic accuracy in the
regime controlled by the quantum critical point:
$q$ scales as $ln \: \omega$.
Formally this corresponds to a dynamical critical exponent
$z_d \rightarrow \infty$.
This appears crucial to understand many of the
observed anomalies in copper-oxide metals.
It should be noted that the propagator (5.10)
for $G=0$ is not the $z_d = \infty$ limit of the
propagators discussed for example in Refs. (13),
which are $\sim (i \omega /q^{\beta} + q^{\alpha} )^{-1}$ with $z_d$
defined to be ($\alpha + \beta$).
\setcounter{equation}{0}
\section{Physical Properties}
The transport properties in regime I (see fig. 7)  which are controlled
by the quantum critical point
and the cross-over to the customary behavior at low temperature for
the overdoped case, region III,
are calculated below.

In the pure limit, region II should
show Fermi-liquid properties but with different
parameters from region III due to the alteration
of states near the Fermi-surface by $T_y \neq 0$.
In the next section I argue that the transition between region I
and II is only a cross-over and that at low temperatures
region II is dominated by disorder such that the density of states
at the chemical potential is zero.
One should however expect a bump in $C_v /T$ and $\chi$
at the I to II crossover.
There is a decrease in low energy fluctuations in region II as $G_0$
in Eq. (5.10) is finite.
But with the density of states at the chemical potential
tending to zero
at low temperatures due to disorder, a Fermi-liquid behavior may
never be observable except in very pure samples.

In the pure limit, although an order parameter develops in
region II, it is by no means clear that there exist observable
singularities in $C_v/T$ or $\chi$ at the transition.
Certainly at $\nu =0$, $z = \infty$, no singularities exist.
The crossover to Ising singularities (only logarithmic
in the specific heat) at high temperatures for $x$ far from $x_c$
may occur with a very small amplitude at observable temperatures.
This requires further work.

\subsection{Single Particle Spectra: 
Angle Resolved Photoemission Experiments \
(ARPES) and Single Particle Tunneling.}

A one-particle self-energy of the form (1.1) was suggested on phenomenological
grounds.$^{12}$
While ARPES were soon found consistent$^{15}$ with this behavior, there has
been since then a considerable development in such experiments.
Closer bounds should be put to this prediction.
A useful formula to fit the self-energy which interpolates
properly between $\omega / T \ll 1$ and $\omega / T \gg 1$,
while obeying analyticity requirements is given by Eq. (6.4).
At low energies and low temperatures this behavior is modified in an
interesting way by defects, as discussed later.

Single particle tunneling has traditionally been a powerful tool for
measuring the frequency dependence of the single particle spectra.
Here, the interpretation of the tunneling spectra is complicated
by the fact that if the self-energy is momentum independent its effect
is not felt in the tunneling spectra unless the tunneling matrix element
is momentum dependent.
The situation has been amply discussed in Ref. (14) and
need not be repeated.
Under suitable conditions,
the conductance as a function of voltage,
$G(V)-G(0) \sim \; Im \; \Sigma (V) \sim V$
as observed for tunneling in the c-direction.
The observed $G(V)$ varies weakly for tunneling in the a-b plane.
This has also been discussed.
A new experimental development is the observation$^{56}$ by inelastic scanning
tunneling microscopy that $[ G(V) - G(0) ] / |V|$
increases as the distance of the tip to the surface increases
thereby decreasing $G(0)$.
This is in accord with  Ref. (14).

As in the case of superconductivity through electron-phonon scattering,
tunneling spectroscopy should serve to identify the spectra of the
glue for superconductivity.$^{57}$
If the collective mode $D(q, \omega )$ is the glue,
the tunneling conductance in the superconducting state, $G(V)-G(0) \sim |V|$ above the superconducting gap as
observed in appropriate geometry.
Quantitative verification of these ideas has been difficult
because the slope of the conductance curve depends on
``extraneous'' factors as discussed.
But it should be possible in carefully designed experiments to
normalize away the extraneous factors.
After normalization the slope should depend only on the
coupling constant which determines the superconducting transition
temperature $T_c$.
\subsection{Long Wavelength Transport Properties}
As discussed in Sec. 1, an enormous constraint is put on a
theory of copper-oxide metals by the fact that if the long wavelength
transport properties are interpreted by kinetic theory or by
semiclassical Boltzmann equations, the scattering rate for momentum
loss measured in electrical resistivity and the scattering
rate for energy loss measured in thermal conductivity have the same
temperature dependence.
Within experimental uncertainty, the single particle scattering rate measured
in tunneling or ARPES also has the same temperature dependence.
This is especially surprising in a theory in which the breakdown of
Fermi-liquid theory is sought through a critical
point where the long wavelength susceptibility diverges.
One might imagine then that only long wavelength fluctuations or
forward scatterings are important in scattering the Fermions, so that
the backward scattering required in momentum transport would make
the transport rate for momentum have a higher temperature dependence than
in the energy transport.

The backward scattering is enforced in momentum transport usually
through considering the two processes shown in (12).
If for instance the Bosons are acoustic phonons, the leading $T^3$ contributions
to the d.c. resistivity of each of the processes in fig. (13) is
exactly cancelled leading to a resistivity proportional to $T^5$.

The situation is quite different with $D(q, \omega )$ of the form (5.10).
Then there is no cancellation to leading order
between the self-energy and the vertex diagrams of fig. (12)
because for energy transfer of order temperature, momentum
transfer throughout the zone is important.
If $D(q,\omega )$ were truly independent of momentum,
as assumed in the marginal Fermi-liquid phenomenology,$^{12}$
(12b) would be identically zero due to the vector nature of the
incoming and outgoing vertices.
This is generally true for any ``s-wave'' scattering.
With $D(q,\omega )$ of the form (5.10), the s-wave
scattering part for $\omega \approx T$ and
$k$ and $k^{\prime}$
both on the Fermi-surface
\begin{equation}
\sim \frac{1}{2 \pi} \int_0^{2 \pi} d \theta \;
\frac{sin^4 \theta}{1+ sin^4 \theta}
\end{equation}
is O(1).
One can show by an explicit calculation that the part of the
process fig. (12b) for $\omega \ll T$ does not change the argument.
The conductivity can therefore be calculated from figs. (12a) alone
with just a numerical renormalization of the coefficient.
Note that due to lattice effects, conservation of momentum with
initial and final states on the Fermi-surface does not imply
conservation of current.
If for arbitrarily small $\omega$, scattering occurs
from a given state on the Fermi-surface to a
substantial part of the Fermi-surface resistivity limited only by
the density of fluctuation results.
Since the {\it density of states} of the fluctuations is essentially
a constant, a linear in $T$ resistivity is to be expected.

The calculation of electrical resistivity, optical
conductivity, thermal conductivity and Raman scattering intensity from (6.1)
is therefore essentially the same as done earlier,$^{12a,b}$ 
with similar results apart from logarithmic corrections. \\
\noindent
{\bf (i)  Optical Conductivity} \\
Optical conductivity as a function of $\omega$ and $T$ is a
much more stringent test of the theory than $\rho (T)$ alone.
Such detailed comparisons with experiments have recently been done
by Abrahams.$^{58}$
Earlier calculations were reported in Ref. (12b).
For completeness and to show the quality of the fit to the
experiments, the experimental results for the inplane
conductivity deduced for untwinned single crystal is shown on
the same scale with the calculations with indicated
parameters in figs. (13a) and (13b), respectively.

The microscopic theory from the strong-coupling limit, Sec. (3)
provides an additional important feature:
The (intraband) optical conductivity sum rule is
\begin{equation}
\int_0^{\infty}  \: \sigma ( \omega ) \: d\omega =
\omega_p^2
\end{equation}
Given the constraint (3.5) the allowed density fluctuations determining
$\omega_p^2$ are only between the one-hole and the two-hole
states $\phi_i$.
>From (4.26) the density of the two-hole states is $x$.
Therefore $\omega_p^2 \sim x (1-x)$.
A proportionality of (7.2) to $x$ for
$x \lesssim 0.2$ has been noted experimentally.$^{59}$

In regime (III), where the integrated
fluctuation spectra is $\sim \omega / G$, a
cross-over from $\rho (T) \sim T$ to $T^2 / G$
and a corresponding change in $\sigma (\omega )$ is
predicted below $T \sim G$.
In the presence of impurities $D(q, \omega )$ becomes optically active.
The mid-infrared bump in the underdoped regime may be attributed to
such processes.
However, in compounds with chains, the chain conductivity$^{60}$ which
appears to have a bump$^{~~}$
as well must be separated for meaningful comparison.

\noindent
{\bf (ii) Thermal Conductivity} \\
The graphs (13) with energy current external vertices
give the thermal conductivity $\kappa (T)$.
It follows the usual kinetic theory expression
\begin{equation}
\kappa (T) \approx \frac{1}{3}
C_v (T) \langle v_F^2 \rangle
\tau_{th} (T)
\end{equation}
with $\tau_{th}^{-1} (T) = \lambda_{th} T$ and
$C_v(T) \approx T \; ln \; \frac{\omega_c}{T}$.
$\lambda_{th}$ departs from $\lambda_{mom}$ by numerical factors
due to the different angular averages in momentum and thermal transport.
The Wiedemann Franz ratio $\kappa (T) / T \sigma (T)$ is
expected to be
$\sim \frac{\lambda_{th}}{\lambda_{mom}} \; ln \; (\omega_c /T)$.

\noindent
{\bf (iii) Raman Scattering Intensity} \\
\noindent
As has been discussed before the Raman intensity in a lattice has a
part proportional to the current-current correlation function and hence
\begin{equation}
S_R (\omega , T) \sim (n (\omega /T ) + 1)
\omega \sigma ( \omega , T)
\end{equation}
So a Raman intensity independent of frequency and temperature near
the ideal composition is expected as observed.
Cross-over to a behavior linear in $\omega$ at low $\omega$ in regimes
II and III is predicted and has been observed in regime (II)
with a cross-over to regime (I) under pressure.
In principle such $S_R(\omega , T)$ is expected in all
polarizations, the relative intensity may in general be quite different.

The collective fluctuations (5.10) also couple directly in the Raman experiment.
But since $Im D (0,\omega ) \sim \omega \sigma ( \omega , T)$,
this also gives the behavior of Eq. (7.4).
\subsection{NMR and Inelastic Magnetic Neutron Scattering}
The application of the theory to the NMR properties has been also
described elsewhere,$^{61}$ so only the principal point is summarized here.
The current fluctuations (5.10) generate an orbital magnetic field which
vanishes at $q=0$ both at the copper site and the oxygen site
as may be seen from fig. 10.
At finite $q$, an orbital magnetic field proportional to $q$ is
generated at the copper sites but not at the oxygen sites.
This is because around the four-fold co-ordinated
Cu site a circulating current due to the electrons
can be constructed to $O(q)$, but not at the two-fold co-ordinated
oxygen site.
This gives rise to an anomalous orbital contribution to
the magnetic correlation functions at the Cu-sites
\begin{equation}
Im \: \chi_{orb} (q,\omega ) \approx
\mu_B^2 (qa)^2 \left ( \frac{a_0}{a} \right )^6
Im D(q, \omega )
\end{equation}
where $a_0$ is the radius of the Cu d-orbital.
The nuclear-relaxation rate calculated using (7.5) has the correct
temperature dependence to fit the observations on Cu.$^{57}$
Oxygen nuclear relaxation rate follows the Koringa law.

One of the most important aspect of the experimental results$^{24,29}$  is
that the oxygen relaxation rate divided by the oxygen Knight shift
does not vary either with $x$ or from compound to compound within
experimental uncertainty.
Given this fact, and the fact that antiferromagnetic fluctuations,
to the extent they are seen, change the position of their peak
and their width with $x$, it is impossible to take seriously
proposals which rely on the cancellation of such fluctuations
at oxygen sites to account for the observations.
A more robust symmetry is called for.
In the picture presented here lattice symmetry never allows
$\chi_{orb} (q, \omega )$ at oxygen sites.

The predicted $\chi_{orb} (q, \omega )$, Eq. (7.5), can be
measured by inelastic neutron scattering.
Perhaps inelastic x-ray scattering can help distinguish the
orbital magnetic fluctuations from spin fluctuations.
Equation (7.5) predicts an unusually smooth $q$-dependence and scattering
up to the high-energy cutoff at any $q$.
The measured $q$-integrated magnetic fluctuation spectrum in
$La_{1.85} Sr_{.15} CuO_4$ is consistent$^{21}$ with (7.5).
Further tests are suggested especially in compounds where nesting features
of the bandstructure do not introduce sharp $q$-dependent features at low
energies.
A direct test of the theory would be
inelastic neutron scattering experiments in several Brillouin zones
and transformation back to real space to deduce separately the magnetic
fluctuations on oxygen and on Cu.
Only a Fermi-liquid contribution
$\sim N(0) \omega / qv_F$ for $\omega \lesssim qv_F$ and 0
beyond should be seen on oxygen, which in appropriate range is negligible
compared to (7.5), which should be seen only on Cu.
\setcounter{equation}{0}
\section{Effects of Impurities}
The problem of disorder in a non-Fermi-liquid is complicated (and interesting).
Only a preliminary treatment of some ideas is presented here to clarify aspects of
the phase-diagram of copper-oxides in the underdoped regime and the fate of the
transition to the circulating current phase in the 
presence of disorder.  

First consider the effect of disorder in the transition from phase I to phase II,
the circulating current phase.
The Harris criteria$^{62}$ may be used in the
classical regime of the transition to determine if quenched disorder, which varies
the {\it local} transition temperature $T_c (r)$ is {\it relevant}.  This is derived
by equating the free-energy contribution due to fluctuation in $T_c (r)$ in a
correlation volume to the pure fluctuation energy in the same volume.  If
\begin{equation}
d \nu - 2 < 0
\end{equation}
disorder is relevant.
Here $\nu$ is defined in terms of the correlation length as
$\xi \sim (T - T_c )^{- \nu}$ for a fixed $x$.
In the Gaussian fluctuation regime, $\nu = 1/2$ while
in the critical fluctuation regime
for the $d=2$ Ising model
$\nu = 1$, so disorder is
relevant in the former and marginally irrelevant in the latter.
This is expected to be true at asymptotically high temperatures far away from
$x = x_c (0)$ in the phase diagram, fig. (12).

Now consider the transition at $T = 0$ as a function of
$( x - x_c (0) )$.  At $ T = 0$ only zero frequency fluctuations come to play.  So
the dynamical critical exponent $z_d$ cannot affect the relevance of disorder.  The
Harris criteria may be expected to therefore to be valid, but we should define $\nu$
through $( \nu \equiv \nu_0 )$, $\xi \sim ( x - x_c (0) )^{- \nu_0}$.  Using (5.10) and
noting that the argument of the logarithmic at $\omega = 0$, $T = 0$
$log |x-x_c (0) |$ scales as $q^2$,
$\nu_0 = 0$.
The Harris criteria then suggests that disorder is
strongly relevant.  Not much definite appears to be known
about the physical state when this is the case.
The best guess is that the phase transition turns into a cross-over and
that a glassy low temperature phase results with random local orientations of the
order parameter.  This is quite reasonable when correlation lengths are short
$( \nu_0 = 0 )$; there is local ordering around each defect with no correlations
building up between regions around different defects.

To summarize the above,
the correlation length exponent changes from 0 to its classical Ising value at
asymptotically high temperatures
and large $x_c(0)-x$.
Correspondingly one expects only a cross-over in
the $x - T$ plane to a glassy circulating current phase.

The problem is even more interesting because the single particle excitations
begin to acquire more singular self-energy than (1.1) or (6.4) 
due to defects.  

It was conjectured$^8$ that a non-Fermi liquid is an insulator for
arbitrarily small disorder (resistivity $\rightarrow \infty$ as
$T \rightarrow 0$) if superconductivity does not intervene at a
higher temperature.  In recent calculations this conjecture has
been supported by some systematic calculations.$^{63}$  The
result of these calculations is that the impurity contribution to
the resistivity in a marginal Fermi-liquid is proportional to
$\ell n T$ below a cross-over temperature
\begin{equation}
T_x \approx ( \omega_c / \pi ) exp ( - \lambda^{-1} 
\sqrt{k_F \ell_o} )  >
\end{equation}
Here $\ell_o$ is the mean free path due to impurities calculated
in the Born approximation.  Below such an energy scale the density
of states at the chemical potential also tends to zero, as 
$( \ell n \omega )^{-1}$.

The observed low temperature resistivity in dirty samples, or in
samples in which superconductivity is suppressed by a large
magnetic field, (as well as the temperature dependence of the
anisotropy in the resistivity) are consistent with these
calculations.$^{64}$   These calculations, relay on using the
marginal self-energy to calculate the impurity scattering vertex
through a Ward-identity.  The experiment gives a $\ell n T$
resistivity in a wide range of doping in underdoped samples, where
in the pure limit the fluctuations have a gap and a marginal
self-energy is not expected.  One possible way this can happen is
if the fluctuations acquire a finite low energy spectral weight
due to disorder.  This would be consistent with excitations in the
glassy state conjectured above.
\section{Superconductive Instability}
\setcounter{equation}{0}
It is only natural that the fluctuations responsible for the
anomalous normal state also lead to the instability to superconductivity.
We again look to the low-energy Hamiltonian, Eq. (6.1),
to deduce the effective interaction in the particle-particle channel.
As usual, this gives for total momentum of the pair equal to zero:
\begin{equation}
\begin{array}{rl}
H_{pair} = \aSU_{k,k^{\prime}} g(k,k^{\prime} ) & g^* (-k,-k^{\prime} )
D(k-k^{\prime}, \omega ) \\ \nonumber
 & \alpha_{-k\sigma^{\prime \prime \prime}}^+
\alpha_{-k^{\prime}\sigma^{\prime \prime}}^+
\alpha_{-k\sigma^{\prime}}
\alpha_{k\sigma} .
\end{array}
\end{equation}
Equation (9.1) is now used to deduce the symmetry channel with the largest
pairing interaction.
The procedure followed is the generalization to more than one
atom per unit cell case of that in Ref. (65),
where it was shown that antiferromagnetic fluctuations promote
even-parity
spin-singlet pairing which has ``d-wave'' symmetry in metals with
appropriate band structures.
The situation here is much more complicated; a
preliminary analysis is given below.
The propagator $D(q,\omega )$ is to a very good approximation
independent of momentum for frequencies $\omega$ of
importance for pairing which are always higher than $T$.
So $D$ can be regarded as a constant, $D_0$ with an upper
frequency cut off $\omega_c$.
The effective pairing Kernel is then
\begin{equation}
\sim \left (
\begin{array}{r}
3 \\
-1
\end{array}
\right )
g_0^2
D_0 | {\cal F}_y (k) + {\cal F}_y (k^{\prime} |^2
sin^2
\left ( \frac{\lambda_k - \lambda_{k^{\prime}}}{2} \right )
\end{equation}
where the upper case corresponds to spin-singlet (even parity) and
lower to spin-triplet (odd parity) pairing.
We wish to express this as a sum over products
of functions of $k$ and $k^{\prime}$ which have the symmetry of the
lattice and which are mutually orthogonal.
Sticking to the lowest lattice harmonics, we look for coefficients in
the expansion
\begin{eqnarray}
(3) & \left [ J_S + J_A A(k) A(k^{\prime} ) +
J_D D(k) D(k^{\prime} ) + ....  \right ] , \\
(-1) & \left [ J_t (T_{kx} T_{k^{\prime}x} +
T_{ky} T_{k^{\prime}y} ) + .... \right ] ,
\end{eqnarray}
where $J_S$ is the coefficient for the simple s-wave
pairing, $J_A$ for pairing of ``extended s-wave form'', i.e.
\begin{equation}
A(k) = cos k_x a+ cos k_y a
\end{equation}
$J_D$ for pairing of ``d-wave form'', i.e.
\begin{equation}
D(k) = cos k_x a - cos k_y a
\end{equation}
and $J_t$ for the odd parity form $sin k_x a$ or
$sin k_ya$.
The ......... in (9.3-9.4) refer to higher lattice harmonics, i.e.
periodic functions of $2 \; k_x a$, $2 \: k_ya$
and so on which we ignore and which are automatically mixed in to the
gap function below $T_c$ due to the nonlinearity in the
gap equation.
$sin^2 \left ( \frac{\lambda_k- \lambda_{k^{\prime}}}{2} \right )$
in (9.2) is a very complicated function, but
it has two properties which help write down the leading dependence
on $k, k^{\prime}$ consistent with lattice symmetry.
It is zero for $k=k^{\prime}$ and peaks when the difference
momenta is maximum possible, i.e. at
($k_x = k_x^{\prime} = \frac{\pi}{a}$,
$k_y - k_y^{\prime} = \pi / a$.
The lowest lattice harmonic satisfying these conditions is
\begin{equation}
1- \frac{1}{2} \left [ cos (k_x-k_x^{\prime} )
a+ cos (k_y - k_y^{\prime} ) a \right ] .
\end{equation}
In $| {\cal F}_y (k) + {\cal F}_y (k^{\prime} ) |^2$,
the only part I keep is a constant; the others give harmonics.
One then gets the relative magnitudes in
units of $g_0^2 D_0 \approx \lambda N^{-1} (0)$
\[
\begin{array}{lll}
\mbox{s-wave~~pairing:} & \; \; & 3 \: J_S =3 \\
\mbox{D-wave~ pairing:} & \; \; & 3 \: J_D = -3/2 \\
\mbox{extend~s-wave ~ pairing:} & \; \; & 3 \: J_A = -3/2 \\
\mbox{triplet ~ pairing:} & \; \; & -1 \: J_t = +1 .
\end{array}
\]
This immediately implies that simple s-wave pairing and triplet
pairing Kernels are repulsive.
In the present case simple s-wave pairing is disallowed simply from the
face that the effective interaction vanishes at long wave length,
the triplet is disallowed because the fluctuations conserve spin.
The Kernel for D-wave and extended s-wave
are attractive and of equal magnitude.
The situation is thus identical to the case of antiferromagnetic fluctuations
with fluctuations peaking at $(k_xa= \pi$, $k_ya= \pi$).

The $T_c$ is determined, as usual from the linear gap equation
projected to the lattice harmonics.
$T_c^D$ for D-wave is in general different than $T_c^A$,
depending purely on the bandstructure, and the chemical potential,
exactly as for the case of antiferromagnetic fluctuations.
For that case and with Cu-O bandstructure on a square lattice d-wave
is formed to be favored in explicit calculations.
The large density of states due to the proximity
to van Hove singularities in the ($\pi , \pi$) direction
favors d-wave.
The same is therefore expected in the present case.
Just as for antiferromagnetic fluctuations, variations in the
bandstructure near the Fermi-surface gives superconducting states
of different symmetries for the same interaction vertex.

The upper cut-off $\omega_c$ of $D(q,\omega )$ is of $O(E_F)$; from
fit to normal state transport experiments, the coupling
$\lambda \approx 0.5$.
As for normal state transport, vertex corrections are
unimportant for calculations of $T_c$ etc.
So a consistent theory can be built.
It should be noted that when $D(q, \omega )$ acquires a gap or
the material has a propensity to insulating behavior, $T_c$ must
go down.
$T_c$ is therefore maximum near $x= x_c(0)$.

It is worth noting that a signature of the glue for
superconductivity is provided by the tunneling conductance.
Under appropriate experimental conditions, as discussed
in Sec. (7a)
and Ref. (14),
$dG(V)/dV$ (above the superconducting gap) is
proportional to the density of state of the glue
for superconductivity weighted by the $q$-dependence of the coupling
constant:
the famous ``$\alpha^2 ( \omega )  F(\omega )$''.$^{57}$
The present theory predicts this to be a constant (with
small corrections) up to the cut-off $\omega_c$.
This is indeed observed but only in some geometries
for reasons discussed in Ref. (14).
Further systematic studies are called for.
Optical conductivity in the superconducting phase
for frequencies larger than twice the gap also can be
used to deduce the glue for superconductivity.$^{66}$
The existing data is again consistent with the form (5.10).

The results of Ref. (65) show that any Bosonic
fluctuations such that the pairing interaction is miminal
for $q=0$ and maximum for $q= (\pi , \pi )$ produces
d-wave pairing.
To distinguish between mechanisms requires
the experiments discussed above.
\setcounter{equation}{0}
\section{Concluding Remarks, Further Theory, Further Experiments}

This investigation has been based on two basic assumptions: 
(i) Breakdown of Landau theory in more than one dimension
requires scale-invariant low energy fluctuations.
(ii) The solid state chemistry of Copper-Oxide is special and responsible for its
special physical properties.  Accordingly, I have formulated the Copper-Oxide model of
Sec. 2 and tried to investigate its 
properties in a systematic manner to find unusual singular
low energy fluctuations.  
The model does have an antiferromagnetic instability at small
doping.  It probably has other finite $q$ instabilities at larger $x$ for a range of
parameters, especially if there is a nesting of the Fermi-surface.  Given the
experimental data,
I do not regard the singular fluctuations near such instabilities
as a solution to the fundamental problems stated in Sec. 1.
I have found a $q = 0$ transition to an unusual Circulating Current Phase on a
line the $x - T$ plane in the general model in the pure limit terminating at a quantum
critical point at $x = x_c$, $T = 0$.  The model has unusual low energy fluctuations in
which the logarithm of the frequency scales with the momentum, so that the fluctuations
are essentially local in space.  
Such local fluctuations are essential
to understand the peculiar normal state transport anomalies in which the momentum
scattering rate, energy scattering rate and the single particle scattering rate are
all proportional to $T$.  They have the right energy scale to give a ``continuous
behavior'' in optical and Raman conductivities from zero frequency to energies of 
O (1 eV).  The current fluctuations also produce local orbital magnetic field fluctuations
which have the symmetry and temperature dependence to account for the extraordinary NMR
relaxation rates on Cu and O nuclei.  The fluctuations also couple to Fermions to give
a superconducting instability with d-wave symmetry favored.
Disorder appears to convert the transition line to the circulating
current phase to a cross-over line due to the quasi-local nature
of the fluctuations.

There exist several incomplete aspects of the theory presented
here.  While it has been shown conclusively that the circulating
current instability does indeed occur in the model, the phase
diagram in the $T - x$ plane has not been determined.  This
requires an explicit numerical solution of the mean-field
equations (4.27)-(4.29) with an assumed set of reasonable
parameters.  I have relied on Refs. (50-52) and general
analyticity condition for fluctuations near an instability to
present a heuristic derviation of the form of the fluctuation
spectrum, Eqs. (4.35)-(4.36).  A better calculation is desirable.
A detailed treatment of the different regimes of fluctuations and
the effect of disorder is needed.  A complete examination of the
superconductive instability $T_c (x)$ is possible and should be
done.

What are the principal experiments on which this paper has
been silent? First is the question of the very interesting 
magneto-transport anomalies.$^{25,26}$
I have indicated in section 1 that the experimental results
do not appear to show that they have asymptotic low energy-temperature
singularities. But even so, the peculiar sub-leading behavior ought to be calculated. It is true that circulating current fluctuations lead to 
chiral scattering in a perpendicular magnetic field. 
As noted in Ref. (28), the temperature dependence of such chiral
scattering is reflected in the magneto-transport anomalies.
But so far I have not succeeded in formulating 
their effect consistently.  Second, very interesting
changes in angle resolved photoemission spectra$^{67}$ have been 
observed in going from region 1 to region 4 of the phase 
diagram of Fig.(1).  It would be very natural to try to 
associate these with the transition or crossover to the circulating
current phase. The observed changes in the spectra are most pronounced
where the Fermi-surface of the ideally-doped 
samples crosses the $(\pi ,0)-( \pi, \pi )$ direction 
and least pronounced where the Fermi-surface 
crosses the $(0,0) - (\pi ,0)$ direction, i.e. the 
changes have $x^2-y^2$ symmetry. 
As discussed in Sec.(4e) the changes in the one-particle 
spectra in the circulating
current phase also do have $x^2-y^2$ symmetry. This is quite intriguing but
 a calculation of the one-particle spectral function in the circulating current
phase is necessary to draw any conclusions. Third, there are aspects of NMR
experiments, especially the anisotropy in the relaxation rate which are not
explained in Ref.(61). Understanding anisotropy effects in NMR requires a theory
of the coupling of fluctuation between different planes. 
 
Are there experiments left to do after the 
$O (5 \times 10^4 )$ already published to test the
conclusions of this paper?
The answer is yes, but most of them are difficult
experiments.

The most direct and convincing test of the theory would be the
observation of the circulating current phase in the circulating
current phase in the under-doped samples and its evolution as a
function of temperature.  As discussed earlier, long-range order
is unlikely but correlation lengths should be large in very pure
samples.  The current patern in Fig. (6) can be observed by Bragg
scattering of polarized neutrons or polarized x-rays.  I estimate
that with polarized neutrons the spin-flip cross-section in the
circulating current phase at the $(1,1)$ Bragg peak is 
$O (10^{-3})$ the nuclear cross-section.

Another test would be evidence for local magnetic fields
in regions 4 and 2
which are estimated to be O(50 Gauss).
As shown in Fig. (6), the local field
is interstitial; there is no magnetic field either on 
Cu or O lattice sites.  Muon spin
resonance would be a way to look for interstitial 
fields if muons were to sit at the
interstitial sites indicated in Fig. (6).

As noted, the spectrum of current fluctuations at
$q = 0$ as a function of $\omega , T$ and $x$ is directly
observable in Raman scattering.
A direct test of the theory would be 
evidence for the fluctuation spectra of Eq.
(5.10) at large $q$ and the difference in its projection on to the Cu
and O sites obtainable by scattering in several Brillouin-zones
mentioned above.  
In the section on NMR and inelastic neutron scattering, I mentioned that
existing neutron scattering in
$La_{1.85} Sr_{.15} Cu O_4$ is consistent with the
magnetic fluctuations$^{21}$ derived from (5.10).
More detailed tests, especially with
scattering over a large range in momentum and frequency in 
$YBa_2 Cu_3 O_{6.9}$ which shows no nesting related peaks in
q-space, are suggested.

In very pure samples of $YBa_2 Cu_3 O_{6.7}$, and (248) which at stoichiometry behaves
as an underdoped material,
the sliver region 4 between regions 1 and 2 in Fig. (1)
where the resistivity falls below the extrapolation from high temperatures clearly
appears.
The cross-over to the circulating current phase in such
samples should not be two broad.
One should look for signatures of this in
thermodynamic experiments, specific heat
and magnetic susceptibility which should show a bump
near the cross-over and a significant decrease below.
Such samples at low temperatures would be particularly
suitable to look for direct evidence discussed above the
circulating current phase. 

The conjecture about magneto-transport$^{28}$ and the behavior 
of the critical fluctuations in
a magnetic field can be tested by a Raman 
scattering experiment in a magnetic field.
The polarized part of the spectra proportional to the 
magnetic field should acquire
singular low energy, low temperature form.

Some of the other tests of the theory have already been mentioned.
These include:
(i) Improved angle resolved photoemission experiments to
verify Eq. (6.4) and its modifications due to impurities deduced
in Ref. (63).
(ii) Measurement of the electronic specific heat in low $T_c$
copper-oxides (for example the single layer Bi compound with 
$T_c \approx 10K$ near ideal composition) to see the 
$T \ell n T$ contribution to the electronic specific heat.
(iii) Controlled single-particle tunneling experiments to see
the spectrum of the ``glue'' for superconductivity.
It should be mentioned that optical conductivity and Raman scattering
experiments for $\omega > 2 \Delta$ can also be used to deduce
the spectrum of the ``glue".

\section*{Acknowledgements}
I have benefitted through comments and discussions on
this work by G. Kotliar, P. Majumdar, A. Ruckenstein, A. Sengupta and
Q. Si.
As mentioned the Appendix is primarily the work of Q. Si.
He also helped improve Sec. (3) considerably.
Over the years I have benefitted through collaborations and discussions 
on the
copper-oxide problem with E. Abrahams, G. Aeppli,
B. Batlogg, E. I. Blount, T. Giamarchi, Y. Kuroda, P. B. Littlewood,
O. Narikiyo, P. Nozieres, K. Miyake,
S. Schmitt-Rink, C. Sire, A. Sudbo and R. E. Walstedt.
Discussions with B. I. Halperin, D. Huse and S. Sachdev on aspects of critical
phenomena relevant to this work are also gratefully acknowledged.
\clearpage
\setcounter{section}{1}
\section*{Appendix A:  One Electron Bandstructure and
Eigen-vectors}
\setcounter{equation}{0}
\renewcommand{\theequation}{A.\arabic{equation}}
The bandstructure and the eigen-vectors for $H_0$ of Eq. (2.2) for
$t_{pp}/t_{pd} < < 1$ are given here.  For $t_{pd} = 0$, the
bonding ``b" and the anti-bonding ``a" bands have dispersions.
\begin{equation}
\epsilon_{a,b}^o ({\bf k}) = \pm \left( ( \Delta_o / 2 )^2 +
4 t_{pd}^2 s_{xy}^2 (k) \right)^{1/2}
\end{equation}
where
$ s_{xy}^2 ({\bf k}) = sin^2 \left( \frac{k_ya}{2} \right)$ 
and the non-bonding band, ``n" is non-dispersive with energy
$- \Delta_{0} /2$.  The eigenvectors may be specified by the band
annihilation operators in terms of these in the orbitals 
$d_{k \sigma}^+$, $p_{xk \sigma}^+$, $p_{yk \sigma}^+$ as in
Eq. (2.5).

The coefficients in (2.5a-b) for $t_{pp}=0$ (specified by a
superscript $o$) are
\begin{equation}
\begin{array}{rl}
u_{ad}^o (k) & = \left(\frac{\Delta}{2} + \epsilon_a^o (k) \right)
/ N_a (k) \\
u_{ax,y}^o (k) & = 2it_{pd} / N_a (k) \\
u_{bd}^o (k) & = \left(\frac{\Delta}{2} + \epsilon_b^o (k) \right)
/ N_b (k) \\
u_{bx,y}^o (k) & = - 2 it_{pd} / N_b (k)
\end{array}
\end{equation}
where $N_{a,b} (k) = [(\frac{\Delta}{2} + \epsilon_{a,b}^o (k))^2
+ 4t_{pd}^2 s_{x,y}^2 (k) ]^{1/2}$.  The non-bonding orbital has
an energy $- \Delta /2$ and is annihilated by 
\begin{equation}
i (sin (k_y a/2 ) p_{xk} - sin (k_x a/2 ) p_y k) /s_{xy} (k)
\end{equation}

The changes in the coefficients in (2.5a-b) are calculated to
first order in $t_{pp}$ using as a perturbation the $O-O$ hopping
Hamiltonian
\begin{equation}
H_1 = 4 t_{pp} \sum_{k, \sigma} sin \left( \frac{k_xa}{2}
\right) sin \left(\frac{k_ya}{2} \right)
p_{xk \sigma}^+ p_{yk \sigma}^+ \; +\; h.c.
\end{equation}
\begin{equation}
\begin{array}{rl}
u_{ad} (k) & = u_{ad}^o (k) + f_{ab} (k) u_{bd}^o (k) \\
u_{ax} (k) & = s_x (u_{ax}^o (k) - f_{ab}
u_{by}^o (k)) + f_{ac} u_{ay}^o s_y \\
u_{ay} (k) & = s_y (u_{ay}^o (k) - f_{ab}
u_{ax}^o (k) ) - f_{ac} u_{ay}^o s_x \\
\end{array}
\end{equation}

For the coefficients in (2.5b) replace $a \leftrightarrow b$ in
(A.4).  In (A.4) 
\begin{equation}
\begin{array}{rl}
f_{ab} (k) & = \frac{4t_{pp}}{(\epsilon_{bk}^o - \epsilon_{ak}^o)}
\; \frac{sin^2 \left(\frac{k_x a}{2} \right)
sin^2 \left(\frac{k_ya}{2} \right)}{s_{xy}^2 (k)}
\; u_{ax}^o (k) \; u_{by}^o (k) \\
& = f_{ba} (k) \\
f_{ac} (k) & =
\frac{4 t_{pp}}{(\frac{\Delta_0}{2} - \epsilon_{ak}^o)}
\; \frac{sin^2 \left(\frac{k_xa}{2}\right) - sin^2 \left(
\frac{k_ya}{2} \right)}{s_{xy}^2 (k)}
\; sin \frac{k_xa}{2} \; sin \frac{k_ya}{2} \\
& = - f_{bc} (k) \\
\end{array}
\end{equation}

\clearpage
\setcounter{section}{1}
\section*{Appendix B:  Exchange Hamiltonian in $\mbox{\boldmath $\tau$ }$-space}
\setcounter{equation}{0}
\renewcommand{\theequation}{B.\arabic{equation}}
This appendix is primarily the work of Q. Si.
Here the exchange Hamiltonian (3.18) is derived from the strong-coupling
limit where the high energy states described in Sec. (3a) are eliminated
by a canonical transformation.

As usual in such a procedure, we write the Hamiltonian as
\begin{equation}
H = H_{low} + H_{high} + H_{mix}
\end{equation}
where $H_{low}$ contains the low energy states we wish to keep, i.e. states
$d_{1i\sigma}^+ |0>$, $d_{2i\sigma}^+ |0>$.
$H_{high}$ the states we wish to discard and $H_{mix}$ connects states in
$H_{low}$ and $H_{high}$ .
We introduce
a canonical transformation
\begin{equation}
\tilde{H} = e^{iS} H \: e^{-iS}
\end{equation}
such that to linear order in $H_{mix}$, matrix elements
connecting states of $H_{low}$ and $H_{high}$ vanish.
This requires
that $S$ be determined by
\begin{equation}
H_{mix} + i \: [ S, \: H_{low} +
H_{high} ] = 0 .
\end{equation}
The transformed Hamiltonian, to second order in $H_{mix}$ is
\begin{equation}
\tilde{H} = H_0 + i \:
[S \; H_{mix} ] ,
\end{equation}
is the kinetic energy (3.4)
$H_{mix}$, which besides the
allowed part in the low energy space (3.16) mixes states of
$H_{low}$ and $H_{high}$.
We give the results for elements of {\bf M} in the approximation
that all the ``{\it Neglected states}'' specified in Sec. (3a)
except the zero-hole state (with energy 0) are assumed to be infinitely
high compared to the low-energy states:
the two one-hole states at energy $\mp \Delta - \mu$ and the two-hole state
at energy $E_{\phi} = V-2 \mu$:
(No essential difference arises in the more general and messy situation.)
Accordingly, we write the kinetic energy in terms of operators
$d_{1i\sigma}$ and $d_{2i\sigma}$ using (3.6), (3.7) and the first term of (3.17).
\begin{equation}
\begin{array}{rl}
d_{i\sigma}^+ &= \frac{1}{\sqrt{2}} sgn  \sigma \phi_i^+
d_{2i-\sigma} + d_{1i\sigma}^+ \phi_{0i} \\
D_{i\sigma}^+ &= \frac{1}{\sqrt{2}} sgn \sigma \phi_i^+
d_{1i-\sigma} + d_{2i\sigma}^+ \phi_i .
\end{array}
\end{equation}
The bare kinetic energy or $H_{mix}$, Eq. (3.4) is
\begin{equation}
\begin{array}{rl}
H_{mix} = \aSU_{i<j,\sigma}
& t_{ij}^{dD} d_{i\sigma}^+ D_{j\sigma} +
t_{ij}^{dd} d_{i\sigma}^+ d_{j\sigma} +
t_{ij}^{DD} D_{i\sigma}^+ D_{j\sigma} \\
 &+ \; h.c.
\end{array}
\end{equation}
Inserting (A.5)  in (A.6), we solve for $S$ in Eq. (A.3) by taking
matrix elements between the states of $H_{low} + H_{high}$ of known energy.
S is then
inserted in (A.4).
The second term gives the exchange Hamiltonian of the form (3.19).
The part in spin $\mbox{\boldmath $\sigma$} $
space is isotropic because of rotational invariance in
$\mbox{\boldmath $\sigma$} $ space.
The part in ($d_1 -d_2$) space is specified below
\begin{equation}
( \tau_y^i \: \tau_x^i \: \tau_z^i ) \; \mbox{\boldmath M} \;
\left (
\begin{array}{c}
\tau_y^i \\
\tau_x^i \\
\tau_z^i
\end{array}
\right ) ,
\end{equation}
{\bf M} has the form
\begin{equation}
\left (
\begin{array}{ccc}
M_{yy} & 0 & 0 \\
0 & M_{xx} & M_{xz} \\
0 & M_{zx} & M_{zz}
\end{array}
\right ) ,
\end{equation}
which we already discussed is the most general form allowable.

We find that with
$E_{\phi} \pm 2 \Delta \equiv E_{\pm}$
and $E_+^{-1} + E_-^{-1} \equiv \bar{E}^{-1}$
\begin{equation}
\begin{array}{rl}
M_{yy}^{ij} &= -2 \: t_{dd}^{ij} t_{DD}^{ji} / E_{\phi} +
(t_{dD}^{ij} )^2 / \bar{E} \\
M_{xx}^{ij} &= -2 \: t_{dd}^{ij} t_{DD}^{ji} / E_{\phi} -
(t_{dD}^{ij} )^2 / \bar{E} \\
M_{zz}^{ij} &= (t_{dd}^{ij2}  + t_{DD}^{ij2} ) / E_{\phi} -
(t_{dD}^{ij} )^2 / \bar{E} \\
M_{zx}^{ij} = M_{xz}^{ij} &= \frac{1}{2}
t_{dD}^{ij} (t_{dd}^{ji} + t_{DD}^{ji} )
\left (
\frac{1}{E_+} + \frac{1}{E_-}
\right ) .
\end{array}
\end{equation}

A rotation (3.32)-(3.33) about $\tau_y$ which diagonalizes the
kinetic energy to a-b space of Sec. (4),
is used to get the Hamiltonian which on
Fourier transforming
gives (4.2).
For the special case $\Delta =0$, the rotation is by
an angle $\frac{\pi}{4}$.
In that case and if $t_{dd}=t_{DD}=0$,
\begin{equation}
\begin{array}{rl}
{\cal J}_{\perp} &=
t_{dD}^2 / E_{\phi} , \;\;
{\cal J}_{zz} = 0 \\
({\cal J}_{xx} - {\cal J}_{yy} ) &=
-t_{dD}^2 / E_{\phi} \\
{\cal J}_{zx} &= \frac{1}{2} t_{dD}^2 / E_{\phi} .
\end{array}
\end{equation}
Then for $\langle \tau_z \rangle =1$, $A+2B =0$.
For the more general case, a condition on $\Delta$ or $x$
can always be found, so that the condition for a QCP derived in
Sec. (4), i.e. $\pm A+2B+4C=0$ is fulfilled.
\clearpage
\section*{References}
\begin{enumerate}
\item{(a)}
The most complete references to the experimental results are the
Proceedings of the last two Tri-annual International Conferences on Materials and
Mechanisms of Superconductivity: Physica C - Volumes {\bf 185-189} (1991) reprinted,
editors M. Tachiki, Y. Muto and Y. Syono, North Holland (1991); Physica C - Volumes {\bf
235-240} (1994) reprinted, editor P. Wyder, North Holland (1994).\\
(b)
A source of
excellent review articles is the series {\it Physical Properties of High Temperature
Superconductors},
editors D. M. Ginsburg, World Scientific, Singapore, Vol I (1989),
Vol II (1990), Vol. III (1992),
Vol. IV (1994).
\item
For a particularly clear exposition of the Landau's phenomenological theory, see
Chapter I of D. Pines and P. Nozi\`{e}res, {\it Quantum Liquids}, Vol. I, Benjamin, New York
(1964); For a microscopic derivation, see Chapter IV of A. A. Abrikosov, L. P. Gorkov
and I. E. Dzyaloshinksii, {\it Methods of Quantum Field Theory in Statistical Physics},
Prentice Hall (1963), and P. Nozi\`{e}res, {\it Interacting Fermi Systems}, Benjamin, New
York (1964).
\item
S. Martin et al., Phys. Rev. B {\bf 41}, 846 (1990).
\item
J. W. Loram et al., Physica C {\bf 235}, 134 (1994) and preprint.
\item
R. Shankar, Rev. Mod. Phys. {\bf 66}, 129 (1994); C. Castellani, C. DiCastro and
W. Metzner, Phys. Rev. Lett. {\bf 72}, 3161 (1994); A. Houghton and J. B. Marston,
Phys. Rev. {\bf 48}, 7790 (1993),
A. H. Castro Neto and E. Fradkin, Phys. Rev. Letters {\bf 72}, 1393 (1994).
\item
For an alternative point of view that the physics in more than 1 dimension is
similar to 1 dimension, see P. W. Anderson, Phys. Rev. Lett. {\bf 64}, 1839 (190); {\bf
65}, 2306 (1992); P. W. Anderson and D. Khevschenko (preprint).
\item
H. Takagi et al., Phys. Rev. Lett. {\bf 69}, 2975 (1992).
\item
G. Kotliar et al., Europhysics Letters {\bf 15}, 655 (1991).
\item
Y. Fukuzumi et al., Phys. Rev. Lett. {\bf 76}, 684 (1996).
\item
B. Bucher et al., Physica C {\bf 157}, 478 (1989).
\item
T. Zhou, K. Syassen, M. Cardona, J. Karpinski and E. Kaldis, Solid State Comm.
(to be published).
\item
C. M. Varma, P. B. Littlewood, S. Schmitt-Rink, E. Abrahams, and A. E.
Ruckenstein, Phys. Rev. Lett. {\bf 63}, 1996 (1989);\\
(a) P. B. Littlewood and C. M.
Varma, J. App. Phys. {\bf 69}, 4979 (1991).
\item
J. Hertz, Phys. Rev. B {\bf 14}, 1165 (1976); A. V. Chubukov, S. Sachdev and J.
Ye, Phys. Rev. B {\bf 49}, 11919 (1994); S. Sachdev, {\it Proceedings of the 19th IUPAP
Conference on Statistical Physics}, Xiamen, China, edited by B.-L. Hao, World
Scientific, Singapore (1996); A. J. Millis, Phys. Rev. {\bf 48}, 7183 (1993).
\item
P. B. Littlewood and C. M. Varma, Phys. Rev. B {\bf 45}, 12636 (1992).
\item
C. G. Olson et al., Science {\bf 245}, 731 (1989).
\item
T. Timusk and D. B. Tanner in Vol. I (1989) and Vol. III (1992) of Ref. 1(b).
\item
Some review articles are: C. Thomsen in {\it Light Scattering in Solids VI}, edited
by M. Cardona (Springer-Verlag, New York 1991); G. Guntherodt, ibid.;
D. Einzel and R. Hackl, J. Raman Spectroscopy, {\bf 27}, (1996).
\item
C. Uher in Vol. III, page 159-284 of Ref. 1(b) (1992).
\item
For an alternative point of view, see for example, T. Moriya, Y. Takahashi and K.
Ueda, J. Phys. Soc. Japan {\bf 59}, 2905 (1990); P. Monthoux and D. Pines, Phys. Rev. B
{\bf 47}, 6069 (1993). 
\item
J. M. Tranquada et al., Phys. Rev. B. {\bf 46}, 5561 (1992); J. Rossat-Mignod et
al., Physica {\bf 192B}, 109 (1993); L. P. Regnault et al., Physica C {\bf 235}, 59
(1994); B. Keimer et al., (preprint).
\item
T. E. Mason et al., Phys. Rev. Lett. {\bf 71}, 919 (1993); S. M. Hayden et al.,
Phys. Rev. Lett. {\bf 76}, 1344 (1996).
\item
For an alternative point of view see, V. J. Emery and S. A. Kivelson, Physica {\bf
209C}, 597 (1993), and Ref. (23).
\item
C. Castellani, C. DiCastro, and M. Grilli,Phys. Rev. Lett. {\bf 75}, 4650 (1995).
\item
An excellent review article is C. H. Pennington and C. P. Slichter, Vol. III of
Ref. 1(b).
\item
(a) T. R. Chien, Z. Z. Wang and N. P. Ong, Phys. Rev. Lett. {\bf 67}, 2089 (1991);
J. M. Harris, Y. F. Yan and N. P. Ong, Phys. Rev. B.  {\bf 46}, 14293 (1992). (b) H. Y.
Hwang et al., Phys. Rev. Lett. {\bf 72}, 2636 (1994).
\item
J. M. Harris et al., Phys. Rev. Lett. {\bf 75}, 1387 (1995).
\item
W. E. Pickett et al., Rev. Mod. Phys. {\bf 61}, 433 (1989).
\item
G. Kotliar, A. Sengupta and C. M. Varma, Phys. Rev. {\bf 53}, 3573 (1996).
\item
C. Berthier et al., Physica Scripta T {\bf 49}, 131 (1993).
\item
Y. Kuroda and C. M. Varma, Phys. Rev. {\bf 42}, 8619 (1990) and Ref. 12(a).
\item
M. Nuss et al., Phys. Rev. Lett. {\bf 66},
3305 (1991).
\item
See for example, H. Ding et. al., Phys. Rev. Lett. {\bf 74}, 2784
(1995).
\item
D. J. Van Harlingen, Rev. Mod. Phys. {\bf 67},
515 (1995),
W. N. Hardy et al.,
Phys. Rev. Lett {\bf 70}, 3993 (1993);
C. C. Tsuei et al., Phys. Rev. Lett. {\bf 73},
593 (1994).
\item
A. G. Sun et al., Phys. Rev. Lett. {\bf 72}, 2267 (1994).
\item
D. H. Wu et al., Phys. Rev. Lett. {\bf 70}, 85 (1993).
\item
C. M. Varma, S. Schmitt-Rink and E. Abrahams, Solid State Com.. {\bf 62}, 681
(1987), and in {\it Novel Mechanisms of Superconductivity}, V. Kresin and S. Wolf eds.,
Plenum (1987).
A model with both Cu and O orbitals was proposed at the same time by V. Emery,
Phys. Rev. Lett. {\bf 58}, 2794 (1987).  But the ionic interactions played no role in that model.
\item
C. M. Varma in {\it Strongly Interacting Fermions and High Tc Superconductivity},
Les Houches, Session LVI (1991), B. Doucot and J. Zinn-Justin eds., Elsevier (1995).
\item J. Zaanen, G. A. Sawatsky and J. W Allen, Phys. Rev. Lett. {\bf 55}, 418 (1985).
\item A. Sudbo et al., Phys. Rev. {\bf 72}, 2478 (1994); E. B. Stechel et al., Phys.
Rev. B. {\bf 51}, 553 (1995); A. Sandvik and A. Sudbo, preprint.
\item L. V. Keldysh and Yu V. Kopaev, Soviet Phys. Solid. State {\bf 6}, 2219 (1965); J.
des Cloizeaux, J. Phys. Chem. Solids {\bf 26}, 259 (1965); D. Jerome, T. M. Rice and W.
Kohn, Phys. Rev. {\bf 158}, 462 (1967).
\item P. B. Littlewood, C. M. Varma, S. Schmitt-Rink and E. Abrahams, Phys. Rev. Lett.
{\bf 63}, 2602 (1989), P. B. Littlewood, Phys. Rev. B {\bf 42}, 10075 (1990); Y. Bang et
al., Phys. Rev. B {\bf 47}, 3323 (1993); S. Coppersmith and P. B. Littlewood, Phys. Rev.
B {\bf 42}, 3966 (1990); R. Putz et al., Phys. Rev. B {\bf 41}, 853 (1990).
\item M. Grilli et al., Phys. Rev. Lett. {\bf 67}, 259 (1991), Phys. Rev. B {\bf 47},
3331 (1993); J. C. Hickes, A. E. Ruckenstein, and S. Schmitt-Rink, Phys. Rev. B {\bf
45}, 8185 (1992).
\item A preliminary and incomplete treatment 
is in C. M. Varma, Phys. Rev.
Lett. {\bf 75}, 898 (1995).
Specifically, the fluctuations are calculated only in the frozen
Fermi-sea approximation, and the symmetry breaking of the
circulating current phase is not specified.
\item B. I. Halperin and T. M. Rice in {\it Solid State Physics}, Vol. 21, F. Seitz D.
Turnbull and H. Ehrenreich eds., (Academic Press, New York, 1968); H. J. Schulz,
Europhys. Lett. {\bf 4}, 609 (1987); A. A. Nersesyan and G. E. Vachnadze, J. Low Temp.
Phys. {\bf 77}, 293 (1989).
\item I. Affleck and J. B. Marston, Phys. Rev. B. {\bf 37}, 3774 (1988); G. Kotliar
Phys. Rev. B. {\bf 37}, 3664. (1988).
\item
Y. Ohta, T. Tohyama and S. Mackawa,
Phys. Rev. Lett. {\bf 66}, 1228 (1991).
\item
S. E. Barnes, J. Phys. F {\bf 6}, 1375 (1976);
P. Coleman, Phys. Rev. B {\bf 29},
3035 (1984); N. Read and D. M. Newns, J. Phys. C {\bf 16}, 3273 (1983).
These methods
have been used on the Cu-O model with nearest neighbor interactions in Ref. 42.
\item See for instance, A. E. Ruckenstein, P. J. Hirschfeld and J. Appel, Phys. Rev. B
{\bf 36}, 857 (1987).
\item G. Baskaran, Z. Zhou and P. W. Anderson, Solid State Com.. {\bf 63}, 973 (1987).
\item J. Gavoret, P. Nozieres, B. Roulet and M. Combescot, J. Phys. (Paris)
{\bf 30}, 987 (1969); A. E. Ruckenstein and S. Schmitt-Rink, Phys. Rev. B {\bf 35}, 7551
(1987).
For a recent discussion of the effects of recoil, see P. Nozi\`{e}res,
J. Phys. I {\bf 4}, 1275 (1994).
\item G. D. Mahan, Phys. Rev. {\bf 163}, 612 (1967);
P. Nozi\`{e}res and C. DeDominicis,
Phys. Rev. {\bf 178}, 1097 (1969).
\item M. Combescot and P. Nozieres, J. de Physique {\bf 32}, 913 (1971).
\item L. P. Gorkov and G. M. Eliashberg,
Soviet Physics-JETP, {\bf 27}, 328 (1968).
\item
A. Ruckenstein and C. M. Varma,
Physica, C{\bf 185}, 134 (1991).
\item T. Holstein, R. E. Norton and P. Pincus, Phys. Rev. B {\bf 8}, 2647 (1973), M.
Reizer, Phys. Rev. B {\bf 39}, 1602 (1989).
\item
T. Hasegawa, M. Nantoh and K. Kitazawa, Jpn. J. Appl. Phys. {\bf 30},
L276 (1991).
\item
J. R. Schrieffer, D. Scalapino and J. W. Wilkins,
Phys. Rev. {\bf 148}, 263 (1966);
W. L. McMillan and J. M. Rowell in {\it Superconductivity}
Edited by R. D. Parks, Marcel Dekker, New York (1969).
\item E. Abrahams (private communications).
\item See of instance, J. Orenstein et al., Phys. Rev. B {\bf 42}, 6342 (1990).
\item Z. Schlesinger et al., Phys. Rev. Lett. {\bf 65}, 801 (1990).
\item C. M. Varma, Phys. Rev. Lett., {\bf 77}, 3431 (1996).
\item A. B. Harris, J. Phys. C {\bf 7}, 1671 (1974).
\item C. M. Varma, Phys. Rev. Lett. (submitted).
\item Y. Ando et. al.,
Phys. Rev. Lett. {\bf 75},
4662 (1995) and preprint.
S. Martin and C. M. Varma (unpublished);
Analysis of Results in S. Martin et al., Phys. Rev. B {\bf 41},
846 (1990).
\item K. Miyake, S. Schmitt-Rink, C. M. Varma, Phys. Rev. B {\bf 34}, 6554 (1986).
\item See for example, P. B. Allen, Phys. Rev. B {\bf 305} (1971).
\item D. S. Marshall et al., Phys. Rev. Lett. {\bf 76}, 4841
(1996); H. Ding et al., Nature, {\bf 382}, 51 (1996).
\end{enumerate}
\clearpage
\section*{Figure Captions}
\begin{itemize}
\item[Fig. 1.]
Schematic Generic phase diagram of the quasi-two-dimensional Copper-Oxide compounds.
$x$ is the density of holes doped in the planes.  $x_c$ is the "optimum" composition.
The antiferromagnetic phase and the superconducting phases, shown inside solid lines,
occur through phase transitions.  A series of cross-overs shown through dashed lines are
discussed in the text.  The impurity density, as inferred from the extrapolation of the
high temperature resistivity to $T = 0$, i.e. assuming Mattheisen's rule decreases as
$x$ increases.
The size of region 4 decreases with increasing disorder for a given x.
\item[Fig. 2.]
The Zaanen-Sawatzky-Allen (ZSA) phase diagram for 3d-transition metal oxides, slightly
modified and showing the schematic change in the position of the transition metal oxides
going from left to right of the periodic table.  The modification is that one of the
axes is the ionic energy $E_x$ defined through particle-hole spectra.  ZSA used the
charge transfer gap $\Delta$ defined through one-particle spectra.  $E_x < \Delta$ due
to particle-hole interactions.  $E_x / W$ is the more appropriate parameter to
characterize the metal-insulator transition than $\Delta / W$.
\item[Fig. 3.a]
Schematic one hole spectra (measured in photoemission) and one particle-spectra
(measured in inverse photoemission) projected on to Cu states and oxygen states in Cu-O
compounds at $1/2$-filling.
\item[Fig. 3.b]
Schematic one-hole spectra and one-particle spectra for a transition metal oxide far to
the left of Cu say Cr-O, projected on to Cr and to O.
\item[Fig. 4.]
The bonding $b$ and the antibonding band $a$ for the two-dimensional band structure from
one-electron theory in the hole representation with chemical potential $\mu$.  Under the
$q = 0$ transition discussed in the text, identical internal re-arrangements in each unit
cell occur.
So the band structure changes merely to the bands $\beta$ and $\alpha$
shown.
\item[Fig. 5.(a)]
The calculated mean-field free-energy as a function of the real part of interband
order parameter $T_x$ described in the text.
\item[Fig. 5.(b)]
For a fixed $T_x = T_x^0$, the free
energy as a function of the imaginary part of the interband order parameter $T_y$.  
$T_y$ takes the value 0 for $p > 0$ and a finite value below $p < 0$ through a second
order transition $p$ is a parameter, defined in terms of the parameters of the
Hamiltonian and for a given compound can be varied by varying the electron density,
temperature or pressure.
$T_y \neq 0$ corresponds to a circulating current pattern in
the ground state shown in Fig. (10).
\item[Fig. 6.]
The deduced ground state current distribution pattern in the circulating
current phase drawn for four cells.  The $+$ and $-$ signs
indicate
magnetic fields pointing up and down.
\item[Fig. 7.]
The theoretical phase diagram in the pure limit.
The effect of impurities is discussed in the text.
\item[Fig. 8.]
The unit cell of Cu-O compounds in the x-y plane and the minimal orbital set: 
$d_{x^2 - y^2}$ orbital of Cu and a $p_x$ and a $p_y$ orbital of oxygen per unit cell.
\item[Fig. 9.(a)]
Exact representation of the interband susceptibility as a function of energy
$\omega$ and momentum $q$.  The lines are exact one-particle Green's functions and
$\Lambda$ is the complete (reducible) vertex. \\
(b) The interband susceptibility in the
ladder diagram approximation to $\Lambda$ of (a).\\
(c) Elementary self-energy and vertex
corrections neglected in (b).
\item[Fig. 10.(a)]
Interband absorption spectrum near the excitonic threshold in the approximation
that band $b$ is infinitely massive - after Ref. (52). \\
(b) Interband absorption spectrum
near the excitonic ledge with finite hole mass.
The excitonic ledge shifts to lower
energy as the interband interactions increase.
\item[Fig. 11.]
Processes for analysis of the low energy Hamiltonian;  (a), (b), (c) are
processes considered for the Boson self-energy.  (d) for the Fermion self-energy.  (e) is
the lowest order correction to the Boson-Fermion vertex.  (f) Leading self-energy due to
anharmonic interaction between the fluctuations.
\item[Fig. 12.]
Elementary processes for optical conductivity.
\item[Fig. 13.(a)]
The calculated optical conductivity based on the theory
which reproduces results based on phenomenology in Ref. (12a).
The parameters used are $\omega_p \approx 2 \: eV$
$\lambda = 0.5$ and $\omega_c = 1200K$.
A soft cutoff is used.
\item[Fig. 13.(b)]
Experimental results for the optical conductibility in the
basal plane, from Ref. (58).
\end{itemize}
\end{document}